\documentclass[preprint,review,12pt]{elsarticle}

\usepackage{epsfig}
\usepackage{amssymb}
\usepackage{amsthm}
\usepackage{amsmath,amsfonts}

\usepackage{url}
\usepackage{epsfig}
\usepackage{multirow}
\usepackage{multicol}
\usepackage{color}
\usepackage{hyperref}
\hypersetup{colorlinks=true}
\usepackage{caption}
\usepackage{algorithm}%
\usepackage{algorithmicx}%
\usepackage{algpseudocode}%
\usepackage{tabularx}
\usepackage{svg}

\journal{Pattern Recognition}

\begin{document}

\begin{frontmatter}



\title{IBVC: Interpolation-driven B-frame Video Compression}


\author[1,2]{Chenming Xu}
\ead{chenming_xu@bjtu.edu.cn}

\author[1,2]{Meiqin Liu\corref{cor1}}
\ead{mqliu@bjtu.edu.cn}

\author[3]{Chao Yao}
\ead{yaochao@ustb.edu.cn}

\author[4]{Weisi Lin}
\ead{wslin@ntu.edu.sg}

\author[1,2]{Yao Zhao}
\ead{yzhao@bjtu.edu.cn}

\address[1]{Institute of Information Science, Beijing Jiaotong University, Beijing 100044, China}
\address[2]{Beijing Key Laboratory of Advanced Information Science and Network Technology, Beijing 100044, China}
\address[3]{School of Computer and Communication Engineering, University of Science and Technology Beijing, Beijing 100083, China}
\address[4]{School of Computer Science and Engineering, Nanyang Technological University, Singapore 639798, Singapore}

\cortext[cor1]{Corresponding author}

\begin{abstract}
Learned B-frame video compression aims to adopt bi-directional motion estimation and motion compensation (MEMC) coding for middle frame reconstruction. However, previous learned approaches often directly extend neural P-frame codecs to B-frame relying on bi-directional optical-flow estimation or video frame interpolation. They suffer from inaccurate quantized motions and inefficient motion compensation. To address these issues, we propose a simple yet effective structure called Interpolation-driven B-frame Video Compression (IBVC). Our approach only involves two major operations: video frame interpolation and artifact reduction compression. IBVC introduces a bit-rate free MEMC based on interpolation, which avoids optical-flow quantization and additional compression distortions. Later, to reduce duplicate bit-rate consumption and focus on unaligned artifacts, a residual guided masking encoder is deployed to adaptively select the meaningful contexts with interpolated multi-scale dependencies. In addition, a conditional spatio-temporal decoder is proposed to eliminate location errors and artifacts instead of using MEMC coding in other methods. The experimental results on B-frame coding demonstrate that IBVC has significant improvements compared to the relevant state-of-the-art methods. Meanwhile, our approach can save bit rates compared with the random access (RA) configuration of H.266 (VTM). The code will be available at \url{https://github.com/ruhig6/IBVC}.
\end{abstract}




\end{frontmatter}


\section{Introduction}

Video compression aims to achieve high-quality reconstruction while maintaining a high compression ratio with the available transmission and storage requirements. It is a crucial subfield of video restoration~\cite{wang2023Versatile} with similar technologies in tasks like video super-resolution~\cite{chen2024High} and video enhancement~\cite{PATIL2022Dual}. Improved reconstruction is also beneficial for efficient human-machine vision~\cite{sheng2024vnvc}, such as  face~\cite{Liu2021Mutual,Qiao2021Deep} or text~\cite{Shohei2023Deep} recognition. Many studies focus on learned coding solutions, such as low delay end-to-end~\cite{lu2019dvc}, and step-by-step contextual~\cite{li2021deep} P-frame compression. Thanks to these improved methods, the compression performance of neural video codecs surpass that of the best traditional codec H.266/VVC~\cite{bross2021overview}. Based on the success of I/P-frame codecs~\cite{li2021deep}, the recently developed learned B-frame codecs~\cite{yang2022Advancing} leverage bi-directional temporal causality for better compression performance.

\begin{figure}[!t]
\centering
\centerline{\includegraphics[width=0.45\linewidth]{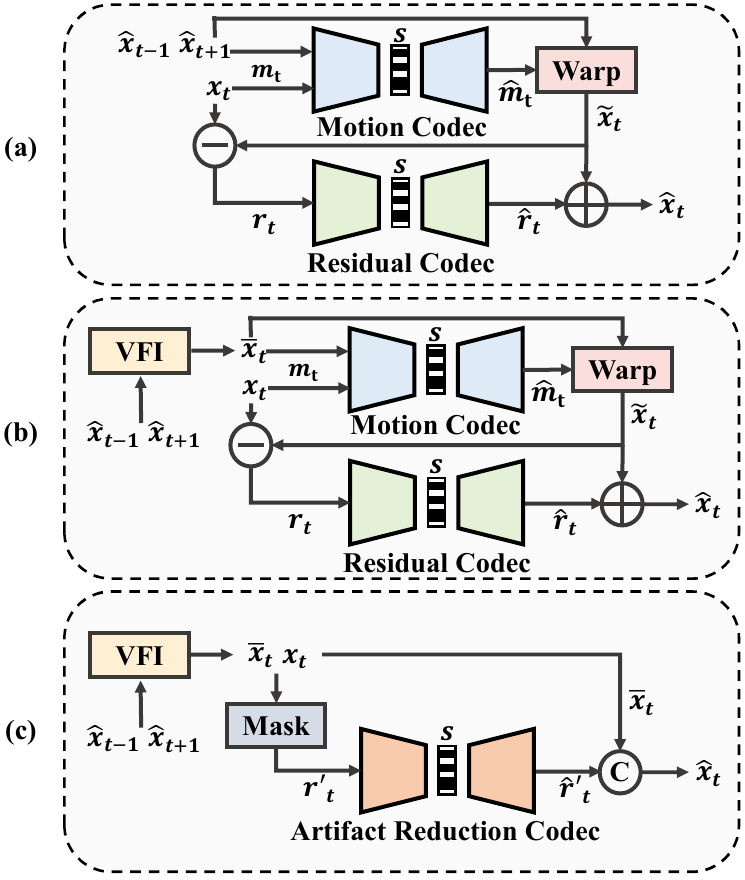}}
\caption{A comparison of different B-frame coding pipelines. \textbf{(a)} B-frame $\hat{x}_t$ prediction is based on bi-directional optical-flow estimation~\cite{yang2020Learning}. The motion vector $m_t$ and the residual $r_t$ are transmitted, respectively. \textbf{(b)} B-frame $\hat{x}_t$ reconstruction is based on video frame interpolation~\cite{pourreza2021extending}. The motion vector $m_t$ and the residual $r_t$ between the interpolated frame $\overline{x}_t$ and the ground truth ${x}_t$ are transmitted. \textbf{(c)} IBVC uses video frame interpolation without excessive bit-rate. The codec only transmits detailed and contextual differences $r'_t$ and reconstructs the multi-scale $\hat{r}'_t$ for artifact reduction.}
\label{fig2}
\end{figure}

As depicted in Figure~\ref{fig2}(a), the prior work~\cite{yang2020Learning} explores a B-frame coding pipeline, where bi-directional motions are coded to achieve better reconstruction. Specifically, the motion $m_t$ between the reference frames $\hat{x}_{t-1}, \hat{x}_{t+1}$ are estimated for transmission. The decoded $\hat{m}_t$ is decoupled to warp the reference frames $\hat{x}_{t-1}, \hat{x}_{t+1}$ for B-frame reconstruction. However, most of these approaches directly extend P-frame coding backbones to the bi-directional setting without considering the quantization distortions on the transmitted $\hat{m}_t$ motion decoupling. The inaccurate motion estimation and compensation also cause artifacts with limited performance improvement. Besides, these methods do not fully leverage bi-directional spatio-temporal dependencies for semantic duplication reduction towards residual coding.

\begin{figure}[!t]
\centering
\centerline{\includegraphics[width=0.55\linewidth]{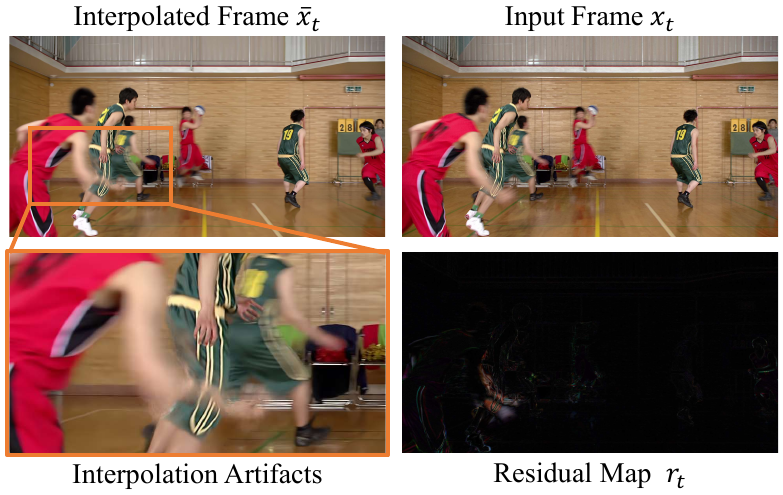}}
\caption{An example from BasketballDrive in HEVC Class B dataset. There are only a few differences between the interpolated frame and the input frame that need to be transmitted. The priority of the codec is to reduce interpolated artifacts by compensating for pixel-level detailed residual.}
\label{fig9}
\end{figure}

It is noted that, as shown in Figure~\ref{fig2}(b), the method~\cite{pourreza2021extending} applies video frame interpolation as pre-alignment to obtain better references for subsequent coding procedures. Specifically, the frame $\overline{x}_t$ is first interpolated by the reference frames $\hat{x}_{t-1}$ and $\hat{x}_{t+1}$. A motion vector $m_t$ between the interpolated frame $\overline{x}_t$ and target frame ${x}_t$ is transmitted to compensate for the location errors~\cite{yang2022Advancing} generated from video frame interpolation. Nevertheless, critical challenges still exist in balancing the performance and the efficiency of B-frame coding. The repetitive MEMC in video frame interpolation and motion codec may lead to excessive computations and annoying artifacts with extra bit rates. In summary, reducing location errors without additional overhead and sufficiently leveraging bi-directional inter-frame information are crucial for B-frame compression.

To tackle these inefficiencies, we propose Interpolation-driven B-frame Video Compression (IBVC). Compared with existing methods, as shown in Figure~\ref{fig2}(c), our strategy deploys a bit-free MEMC using interpolation, reducing the computational redundancy and new compression artifacts. Since the interpolated frame $\overline{x}_t$ has already restored most high-frequency semantics compared to target frame ${x}_t$, as shown in Figure~\ref{fig9}, the codec only needs to fine structures and remove interpolation artifacts. Therefore, our encoder uses a masking strategy to focus on informative regions, guided by the residual between the interpolated frame $\overline{x}_t$ and ground truth ${x}_t$. The proposed decoder restores high-quality frames with artifact reduction by leveraging prior multi-scale intra/inter-frame dependencies. The experiments show that IBVC achieves exceptional reconstruction efficiency and reduces bit-rate requirements compared to state-of-the-art methods. In summary, our main contributions are listed as follows.
\begin{itemize}
\item We design a simple yet effective pipeline for the learned B-frame video codec, which includes an interpolation-driven MEMC and an artifact reduction codec for intermediate frame compression.

\item The proposed residual-guided masking encoder excels at precisely identifying intricate variances and interpolation artifacts, leading to a reduction in redundant bit rates within aligned areas.

\item To rectify interpolation location errors, the conditional spatio-temporal contextual decoder leverages prior conditions from historical frames to enhance temporal consistency and reduce interpolation artifacts.

\end{itemize}

\section{Related Work}

The current B-frame compression methods mainly refer to the learned I-frame and P-frame compression structures. Additionally, there are also some works that utilize video frame interpolation to assist in B-frame compression. We briefly review these related works in the following sections.

\subsection{Learned Image Compression} 
Great progress has been made in the development of learned image compression recently. Ballé \emph{et al.}~\cite{balle2017end} propose variational auto-encoder (VAE) that utilizes factorized and hyperprior entropy models, achieving performance comparable to H.265 (HM) intra-frame coding. The Gaussian mixture model~\cite{cheng2020learned} further improves the entropy codec performance to be on par with H.266 (VTM) intra-frame coding. 

\subsection{Learned P-frame Video Compression}
Inspired by the learned image compression,  Lu \emph{et al.}~\cite{lu2019dvc} and Li \emph{et al.}~\cite{li2021deep} propose a series of end-to-end P-frame video compression methods based on VAE architecture. Sheng \emph{et al.}~\cite{sheng2022temporal} investigate a new propagation strategy that uses the last reconstructed frame and its feature for temporal context mining. Guo \emph{et al.}~\cite{guo2023learning} propose a novel cross-scale prediction module for content-adaptive motion compensation. Later, Li \emph{et al.}~\cite{li2022hybrid} introduce a flexible-rate codec with hybrid spatial-temporal entropy modeling that sets a new state-of-the-art in neural video coding, which surpasses the low delay P-frame mode of H.266 (VTM).

\subsection{Learned B-frame Video Compression}

\subsubsection{Bi-directional Motion Estimation and Compression}
As shown in Figure~\ref{fig2}(a), to further exploit bi-directional correlations, Djelouah \emph{et al.}~\cite{djelouah2019neural} employ bi-directional optical flow compensation for B-frame coding. Yang \emph{et al.}~\cite{yang2020Learning} propose hierarchical learned video compression with three bi-directional warping layers and a recurrent enhancement network. To gain more precise motion, Yılmaz \emph{et al.}~\cite{ccetin2022flexible} exploit rate-distortion optimized and flexible-rate methods to improve the B-frame compression performance. Moreover, Chen \emph{et al.}~\cite{chen2022b} first apply conditional augmented normalization flow (CANF) for bi-directional inter-frame coding.

\subsubsection{{Video Frame Interpolation}}
{Video frame interpolation (VFI) is used to generate new middle frames from existing reference frames without bit-stream. The optical flow~\cite{kong2022ifrnet} and deformable convolution~\cite{ding2021cdfi} are used to do the motion estimation and composition between adjacent frames for intermediate frames reconstruction. For instance, Jiang \emph{et al.}~\cite{jiang2018super} first propose an end-to-end convolutional neural network, named Super-SloMo, designed for variable-length multi-frame video interpolation which can extend for B-frame compression. However, VFI suffers from inaccurate motion estimation and compensation, leading to motion artifacts and prediction misalignment. To address this, video compression can be utilized to further enhance the reconstruction quality.}

\subsubsection{VFI based B-frame Video Compression}
As depicted in Figure~\ref{fig2}(b), some researchers extend interpolation as a historical prior for B-frame coding. Wu \emph{et al.}~\cite{wu2018video} first utilize frame interpolation to optimize the MEMC coding. Pourreza \emph{et al.}~\cite{pourreza2021extending} adopt the Super-SloMo to interpolate the intermediate frames, which are then input as reference frames to P-frame codecs for B-frames coding. Jia \emph{et al.}~\cite{jia2022neighbor} propose a frame synthesis model consisting of two different bit-rate modes for B-frame compression. Yang \emph{et al.}~\cite{yang2022Advancing} utilize the in-loop frame synthesis module to generate a better reference before compression. Alexandre \emph{et al.}~\cite{alexandre2023hierarchical} use two-layer CANF after frame interpolation to replace the motion codec for B-frame video coding. However, these methods do not make full use of prior references, resulting in compression structural redundancy and temporal inconsistency. Additionally, it is challenging to use these methods to address the artifacts caused by quantization in MEMC coding.

\section{Methodology}

\subsection{Motivation}

Computational complexity and compression artifacts are the common issues in recent B-frame coding algorithms, whether traditional codecs~\cite{bross2021overview} or learned-based methods~\cite{yang2020Learning}. Although subsequent methods~\cite{yang2022Advancing} utilize frame interpolation for compression efficiency, two pieces of bit-stream still need to be transmitted for motion and residual compression reconstruction, as:

\begin{equation}
\hat{x}_t = F_{codec}(F_{vfi}(\hat{x}_{t-1},\hat{x}_{t+1}), x_t)
\label{eq1}
\end{equation}
where $x_t$ denotes the original input frame for transmission. $\hat{x}_t$ represents the decoded frame. $\{\hat{x}_{t-1},\hat{x}_{t+1}\}$ are the decoded reference frames. $F_{vfi}(\cdot)$ means the video frame interpolation network, which involves the same MEMC as motion transmission. $F_{codec}(\cdot)$ consists of the MEMC and the residual codecs, where the quantization may result in compression artifacts.

Concretely, in frame interpolation blocks~\cite{kong2022ifrnet}, the middle frame is generated using bi-directional warping, as: 

\begin{equation}
\begin{aligned}
\overline{x}_t &= O \cdot F_{warp}(\hat{x}_{t-1},m_{\hat{x}_{t-1}\rightarrow x_{t}}) \\
&+ (1- O) \cdot F_{warp}(\hat{x}_{t+1}, m_{\hat{x}_{t+1}\rightarrow x_{t}}) 
\label{eq3}
\end{aligned}
\end{equation}
where $\overline{x}_t$ denotes the interpolated frame, $O \in
[0,1]$ means the occlusion coefficient. The non-linear motions $\{m_{\hat{x}_{t-1}\rightarrow x_{t}},m_{\hat{x}_{t+1}\rightarrow x_{t}}\}$ between the middle frame $x_t$ and reference frames $\{\hat{x}_{t-1},\hat{x}_{t+1}\}$ are estimated using the flow distillation learning~\cite{kong2022ifrnet}. $F_{warp}(\cdot)$ denotes the flow warp function. Then, the previous methods~\cite{yang2022Advancing,pourreza2021extending} assume that there may still be unaligned artifacts from location errors~\cite{yang2022Advancing} and align the interpolated frames with the target frames using a motion codec, as:
\begin{equation}
\tilde{x}_t =F_{warp}(\overline{x}_t, F_{dec}(\left\lfloor F_{enc}(m_{\overline{x}_t\rightarrow x_t})\right\rceil))
\label{eq4}
\end{equation}
where $\tilde{x}_t$ is the reconstructed result after flow warping. $F_{enc}(\cdot)$ and $F_{dec}(\cdot)$ mean the encoder and decoder networks. 

\begin{figure}[!t]
\centering{
\includegraphics[width=0.99\linewidth]{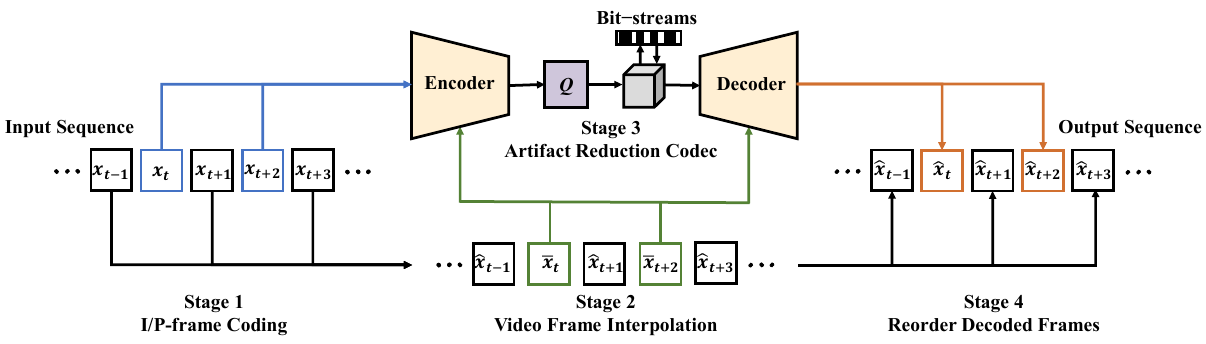}
}
\caption{Overview of proposed pipeline for B-frame video coding. In the design, we use a video sequence consisting of five consecutive frames $\{x_{t-1}, x_{t}, x_{t+1}, x_{t+2}, x_{t+3}\}$ to explain compression process. The pipeline consists of four stages. Firstly, I/P-frame coding is deployed on $\{x_{t-1}, x_{t+1}, x_{t+3}\}$ to generate reference frames $\{\hat{x}_{t-1}, \hat{x}_{t+1}, \hat{x}_{t+3}\}$ for frame interpolation. Later, the interpolated frames $\{\overline{x}_{t}, \overline{x}_{t+2}\}$ and corresponding ground truth frames $\{{x}_{t}, {x}_{t+2}\}$ are fed into the artifact reduction codec to generate reconstructed frames $\{\hat{x}_{t}, \hat{x}_{t+2}\}$. Finally, the video sequence is resorted to its original playing order.}
\label{fig3}
\end{figure}

It is worth noting that there is a major limitation in this approach. Specifically, in each alignment operation, the common target for both flow warp operations in Equation~\ref{eq3} and Equation~\ref{eq4} is the middle frame $x_t$. While this procedure may reduce the location errors between raw frames and ground truth, it also causes notable artifacts due to redundant quantized optical flow in Equation~\ref{eq4}. And MEMC coding also increases the overall computational cost. To address this limitation and eliminate the extra operation in Equation~\ref{eq4}, we directly propagate the aligned frame $\overline{x}_t$ to an artifact reduction codec. The complete coding framework is illustrated in Figure~\ref{fig3} and detailed in Section~\ref{sec2}.

\subsection{Interpolation-driven B-frame Compression Pipeline}
\label{sec2}

\begin{algorithm}[!t]
\caption{{B-frame Codec Algorithm}}\label{algo1}
\begin{algorithmic}[1]
\renewcommand{\algorithmicrequire}{ \textbf{Input:}}     
\renewcommand{\algorithmicensure}{ \textbf{Output:}}    
\Require Original frames $\{x_{t-1},x_{t},x_{t+1},x_{t+2},x_{t+3}\}$.
\Ensure Decoded frames $\{\hat{x}_{t-1},\hat{x}_{t},\hat{x}_{t+1},\hat{x}_{t+2},\hat{x}_{t+3}\}$.

\noindent{// Stage1: Utilize DCVC-HEM for I/P-frames compression.}
\State $\hat{x}_{t-1},\hat{x}_{t+1},\hat{x}_{t+3} =$ DCVC-HEM$(x_{t-1},x_{t+1},x_{t+3})$; 

\noindent{// Stage2: Utilize IFRNet for frame interpolation.}
\State $\overline{x}_t,\overline{x}_{t+2} =$ IFRNet$(\hat{x}_{t-1},\hat{x}_{t+1},\hat{x}_{t+3})$;

\noindent{// Stage3: Utilize artifact reduction codec for B-frame reconstruction.}
\State Feature buffer $ \leftarrow \hat{y}_t,\hat{y}_{t+2} =$ Encoder$(x_{t}|\overline{x}_t,x_{t+2}|\overline{x}_{t+2})$;
\State $\hat{x}_t =$ Decoder$(\hat{y}_{t}|(\hat{y}_{1},...,\hat{y}_{t-2}),\overline{x}_t)$;
\State $\hat{x}_{t+2} =$ Decoder$(\hat{y}_{t+2}|(\hat{y}_{1},...,\hat{y}_{t}),\overline{x}_{t+2})$;

\noindent{// Stage4: Organize decoded frames for output order.}
\State {\textbf{return}} $\{\hat{x}_{t-1},\hat{x}_{t},\hat{x}_{t+1},\hat{x}_{t+2},\hat{x}_{t+3}\}$.
\end{algorithmic}
\end{algorithm}

Figure~\ref{fig3} and Algorithm~\ref{algo1} depicts the proposed coding pipeline for five consecutive frames $\{x_{t-1},x_{t},x_{t+1},x_{t+2},x_{t+3}\}$. The frames are compressed by two main parts: video frame interpolation and artifact reduction codec, as:
\begin{equation}
\hat{x}_t = F_{art}(F_{vfi}(\hat{x}_{t-1},\hat{x}_{t+1}), x_t)
\label{eq16}
\end{equation}
where $F_{art}(\cdot)$ means the artifact reduction codec, which transmits only the detailed contextual differences without repeating the motion transmission.

{As the step 1-2 in Algorithm~\ref{algo1}, the proposed coding pipeline utilizes the decoded I/P-frames $\{\hat{x}_{t-1},\hat{x}_{t+1},\hat{x}_{t+3}\}$ as the input for interpolation using IFRNet-S~\cite{kong2022ifrnet}, which generates preliminary middle frames $\{\overline{x}_t,\overline{x}_{t+2}\}$ in both encoder and decoder.} If $\hat{x}_{t-1}$ is the first frame of a Group of Pictures (GoP), it is compressed using I-frame coding. Otherwise, the sequence frames $\{\hat{x}_{t-1},\hat{x}_{t+1},\hat{x}_{t+3}\}$ are compressed using P-frame coding. The procedure directly implements the MEMC in video frame interpolation $F_{vfi}(\cdot)$ between the target frames and reference frames without any bit-rate consumption. Unlike general MEMC coding methods, our approach does not require quantization of the optical flows for transmission. As a result, it can produce more plausible restored structures with fewer compressed distortions.

As shown in Figure~\ref{fig9}, the interpolated frames have the most reconstructed features compared to input frames. {IBVC is only required to code the difference between interpolated frames $\{\overline{x}_t,\overline{x}_{t+2}\}$ and ground truth frames $\{{x}_t,{x}_{t+2}\}$ by artifact reduction encoder in step 3 of Algorithm~\ref{algo1}}. Since the MEMC process does not consume any bit rate, it is also crucial to leverage the effective inter-frame spatio-temporal information to minimize MEMC errors. {Hence, we consider the triplet $\{\hat{x}_{t-1},{x}_{t},\hat{x}_{t+1}\}$ as a reference unit and utilize the previously encoded representations to be prior conditions during decoder in step 4-5 of Algorithm~\ref{algo1}, as:}
\begin{equation}
 \hat{y}_t = \hat{y}_t | (\hat{y}_{1},..,\hat{y}_{t-2})
\label{eq6}
\end{equation}
where $\{\hat{y}_{1},..,\hat{y}_{t-2}\}$ are the encoded representations in the previous reference units. This strategy forms an intra/inter-frame Markov dependence macroscopically, enabling the reuse of encoded information between frames. More bit rates can be allocated to the artifact areas, compensating for interpolation errors, without introducing excessive bit rate and complexity consumption.

\subsection{Artifact Reduction Encoder-Decoder Network}

\begin{figure}[!t]
\centering{
\includegraphics[width=0.99\textwidth]{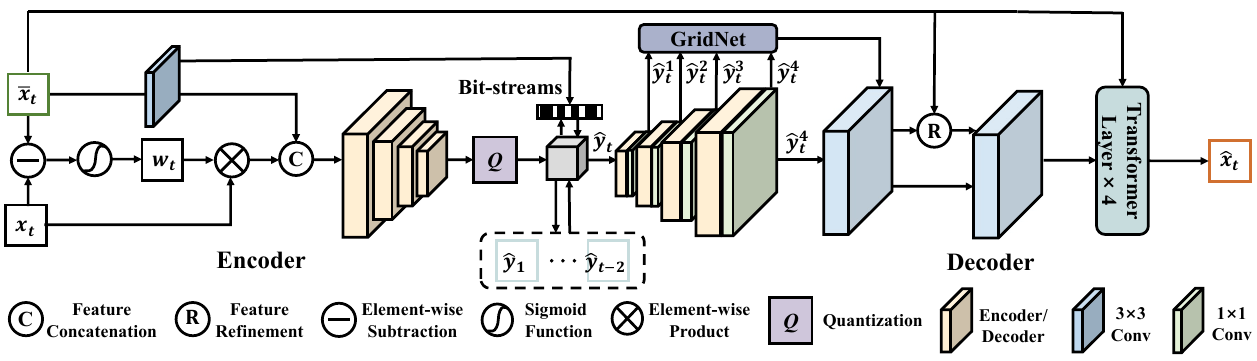}
}
\caption{Illustration of artifact reduction encoder-decoder network. In the design, we propose a residual guided masking encoder to adaptively select informative contexts for productive artifact reduction compression. Later, a conditional spatio-temporal contextual decoder facilitates prior conditions from historical frames to eliminate MEMC artifacts.}
\label{fig4}
\end{figure}

As described in Figure~\ref{fig9}, the codec needs to generate refined visual quality and remove interpolation artifacts, as the interpolated frames have already restored most semantics. However, the existing residual codecs~\cite{lu2019dvc} and conditional codecs~\cite{li2021deep} are oriented towards pixel-level distortions and high-frequency contents, respectively. Moreover, the codec should be required to rely on bit streams for artifact reduction. Therefore, as shown in Figure~\ref{fig4}, we propose a Residual Guided Masking Encoder (RGME) to reduce duplicate coding of aligned areas and provide interpolation conditions. 

Subtle information at the pixel-level is mainly contained in the residual from unalignments between input frames and reference frames. The context provides the bi-directional correlated conditions in feature dimensions from interpolation. To this end, we employ the pixel-level residual information to mark the key contextual region for conditional compression, as:
\begin{equation}
{\hat{y}_t = \left\lfloor F_{enc}(F_{\sigma} (|x_t-\overline{x}_t|) \otimes {x}_t|\overline{x}_t)\right\rceil }
\label{eq7}
\end{equation}
where $\otimes$ means a vector-matrix multiplication operation with dimensionality preserving. The key contextual region is guided by the weight map $w_t = F_{\sigma} (x_t-\overline{x}_t) \in (0,1)$. A sigmoid function $F_{\sigma}(\cdot)$ is used to allocate bit rates towards artifact reduction. $F_{enc}(\cdot)$ means the encoder network. $\left\lfloor\cdot\right\rceil$ means the quantization operation. $\hat{y}_t$ denotes the encoded latent representation by auto-regressive entropy model~\cite{li2021deep}. Unlike the mask guided by motion estimation~\cite{hu2022coarse} or random sampling~\cite{tong2022videomae}, the proposed strategy is specially designed for the interpolation-driven codec. It accurately locates artifacts and transmits the corresponding contextual information, taking advantage of both residual and conditional coding. 

Moreover, to learn intra/inter-frame dependencies from highly compressed latent representations $\hat{y}_t$, a Conditional Spatio-Temporal Contextual Decoder (CSCD) is proposed for multi-scale encoded feature learning and artifact reduction. As illustrated in Figure~\ref{fig4}, our method has adaptively selected the context to encode as highly compressed representations. To incorporate temporal prior conditions from historical frames, the current coded latent representation $\hat{y}_t$ is combined with the previous features. As described in Section~\ref{sec2}, the prior representations achieve the auxiliary effect in MEMC distortion reduction with temporal inter-frame consistency. We utilize the 1$\times$1 convolution and bilinear interpolation to exploit the multi-scale features in the decoder. Then, GridNet~\cite{fourure2017residual} and channel-wise self-attention~\cite{zamir2022restormer} are conducted as a cascaded frame synthesis module to combine the multi-scale spatial features for further artifact reduction, as:
\begin{equation}
 \hat{x}_t = F_{atten}(F_{refine}(F_{grid}(\hat{y}^1_t,...,\hat{y}^l_t), \hat{y}^l_t, \overline{x}_t), \overline{x}_t)
\label{eq11}
\end{equation}
where $\{\hat{y}^1_t,...,\hat{y}^l_t\}$ with $l=4$ denotes the multi-scale features extracted from $\hat{y}_t$. $F_{refine}(\cdot)$ means the spatial context refinement function consists of 3$\times$3 convolutional layers with LeakyReLU activation. $F_{grid}(\cdot)$ means the multi-scale feature extraction. The attention module $F_{atten}(\cdot)$ contains four Transformer layers and is employed to further reduce the artifacts in the latent representations. By utilizing the artifact reduction encoder-decoder network to transmit and explore spatio-temporal correlations, IBVC is able to adaptively restore plausible structures without additional MEMC coding.

\subsection{Objective Function and Training Strategy}
We employ an end-to-end error propagation strategy to reduce training complexity. It integrates a series of rate-distortion (R-D) costs between decoded B-frames and the corresponding ground truth. As a result, a joint objective function is designed to optimize the encoder-decoder network, and defined as the trade-off between bit rates and the reconstruction quality. The compression performance of multiple frames with bi-directional dependencies is also considered in the objective function, as:
\begin{equation}
L = \frac{1}{T}(L_t+L_{t+2}) ~ with ~ L_t = R(\hat{y}_t) + \lambda \cdot D (x_t, \hat{x}_t)
\label{eq13}
\end{equation}
where $L$ means the average R-D cost of $L_t$ with $T$ (set as 2) compressed B-frames. $\lambda$ denotes the trade-off coefficient between the bit rates cost $R$ and the distortion $D$. Following with DCVC~\cite{li2021deep}, we train 4 models for mean squared errors (MSE) with $\lambda = 256, 512, 1024, 2048$, multi-scale structural similarity (MS-SSIM)~\cite{wang2004image} with $\lambda = 8, 16, 32, 64$ and learned perceptual image patch similarity (LPIPS)~\cite{zhang2018unreasonable} with $\lambda = 2, 4, 8, 16$.

\section{Experiments}

\subsection{Implementation Details}
\subsubsection{Training Configuration} For a fair comparison, we use the same training dataset and pre-processing as the previous method DCVC~\cite{li2021deep}. The Vimeo-90K septuplet~\cite{xue2019video} is utilized to train artifact reduction codec, which contains 4,278 videos with 89,800 independent sequences of $256\times256$. The 2nd and 4th frames of each septuplet are the coding targets. The consecutive five frames are used as the input for the interpolation network and codec. All experiments are implemented on the 2 NVIDIA GeForce RTX 3080Ti GPUs with Intel(R) Xeon(R) Gold 6248R CPUs. We conduct a mini-batch size of 4. The training process includes 5 epochs with a learning rate of $1e^{-4}$ and 1 epoch for fine-tuning used $1e^{-5}$. The Adam optimizer is utilized with $\beta_1=0.9$ and $\beta_2=0.999$.

\subsubsection{Testing Configuration} The experimental model is evaluated on the commonly used 1080P UVG dataset~\cite{mercat2020uvg} with 7 sequences. We also utilize the 19 video sequences of HEVC datasets~\cite{bossen2013common} including Class B $(1920 \times 1080)$, Class C $(832 \times 480)$, Class D $(416 \times 240)$, Class E $(1280 \times 720)$ and Class E' $(1280 \times 720)$. We test 96 frames for each video sequence. The GoP size is set as 32. There are 47 reference units with a separate P-frame. Each reference unit has 3 frames. Precisely, 3 I-frames and 46 P-frames are provided by the existing video compression method DCVC-HEM~\cite{li2022hybrid}. The remaining 47 B-frames are compressed by IBVC with interpolated reference frames generated by IFRNet-S~\cite{kong2022ifrnet}. 

\subsection{Evaluation against State-of-the-art Methods}

To verify the effectiveness of IBVC, we make a fair comparison with state-of-the-art learned codecs and traditional codecs in RA configuration. As for H.265/HEVC and H.266/VVC, we compare with the HM-17.0~\cite{hm} and VTM-19.0~\cite{vtm} in RA configuration, respectively. {For a fair comparison on RGB content, the detailed command~\cite{sheng2022temporal} for HM and VTM is shown as follows.}

\begin{itemize}
\item {\textit{EncoderApp -c encoder\_randomaccess\_vtm.cfg -f 96 -q QP --IntraPeriod=32 -i input.yuv --Level=6.2 --InputBitDepth=10 --InputChromaFormat=444 --ChromaFormatIDC=444 --DecodingRefreshType=1 -b output.bin -o input.yuv}}
\end{itemize}

{In addition, the state-of-the-art learned P-frame compression methods, i.e., DCVC~\cite{li2021deep}, DCVC-TCM~\cite{sheng2022temporal} and DCVC-HEM~\cite{li2022hybrid}, the state-of-the-art learned B-frame compression methods, i.e., B-EPIC~\cite{pourreza2021extending}, LHBDC~\cite{yilmaz2021end}, B-CANF~\cite{chen2022b}, ALVC~\cite{yang2022Advancing} and TLZMC~\cite{alexandre2023hierarchical}, are considered for comparison.} For methods without available code, we display the results provided by authors.

\begin{figure*}[!t] 
 \centering
 \begin{minipage}[b]{0.325\linewidth} 
   \centering
   \centerline{\includegraphics[width=\linewidth]{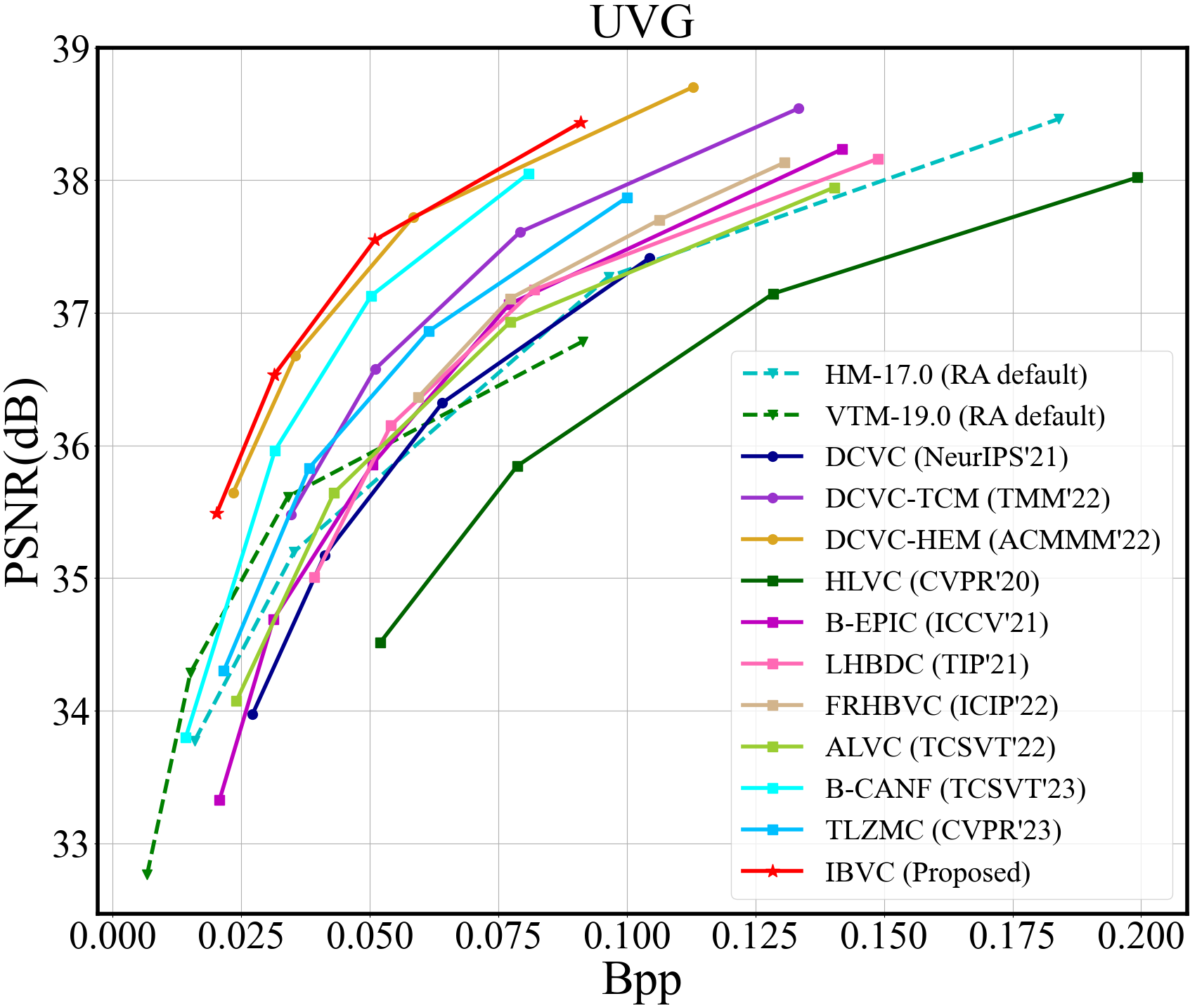}}
 \end{minipage}
 \hfill
 \begin{minipage}[b]{0.325\linewidth}
   \centering
   \centerline{\includegraphics[width=\linewidth]{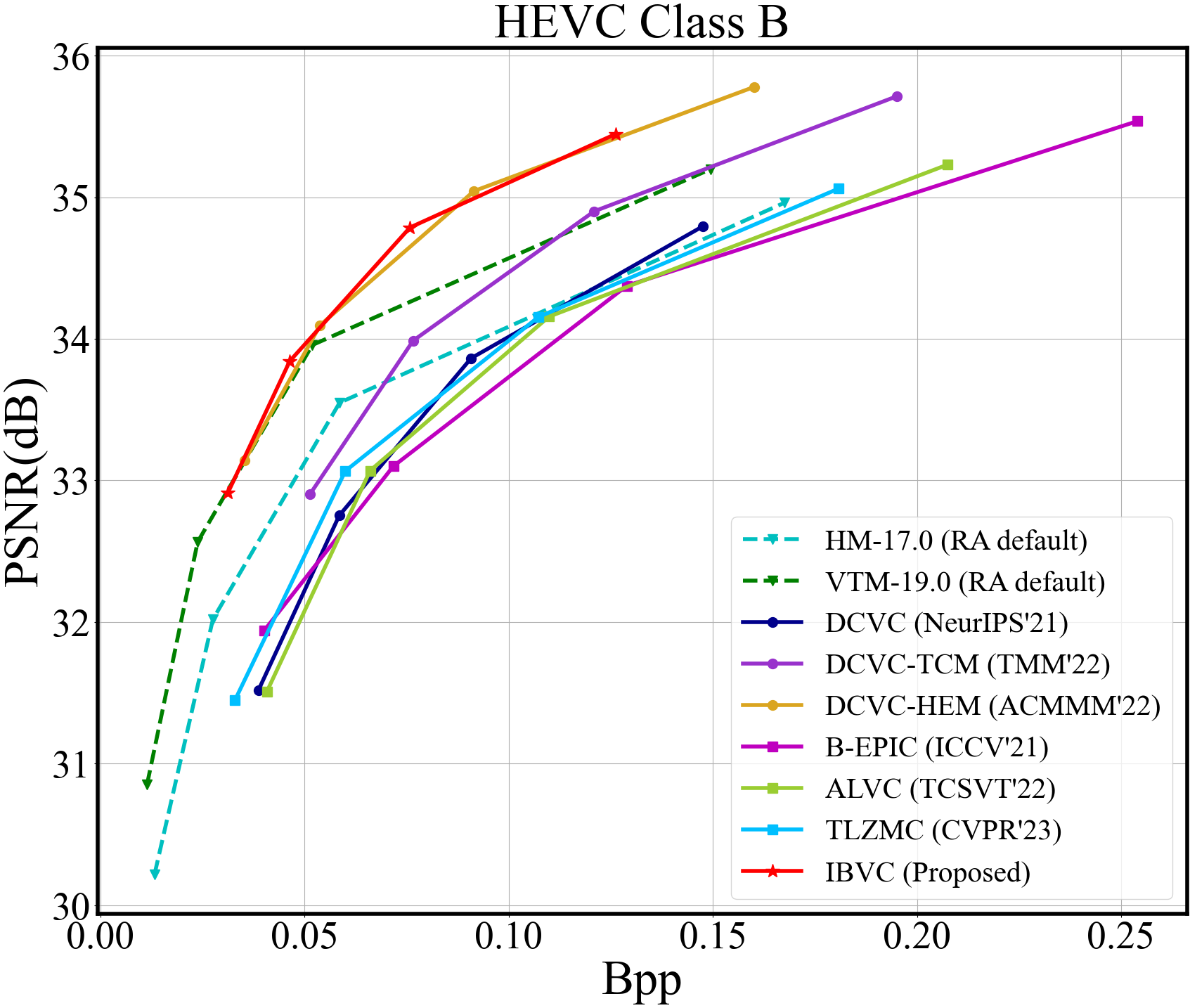}}
 \end{minipage}
 \hfill
 \begin{minipage}[b]{0.325\linewidth}
   \centering
   \centerline{\includegraphics[width=\linewidth]{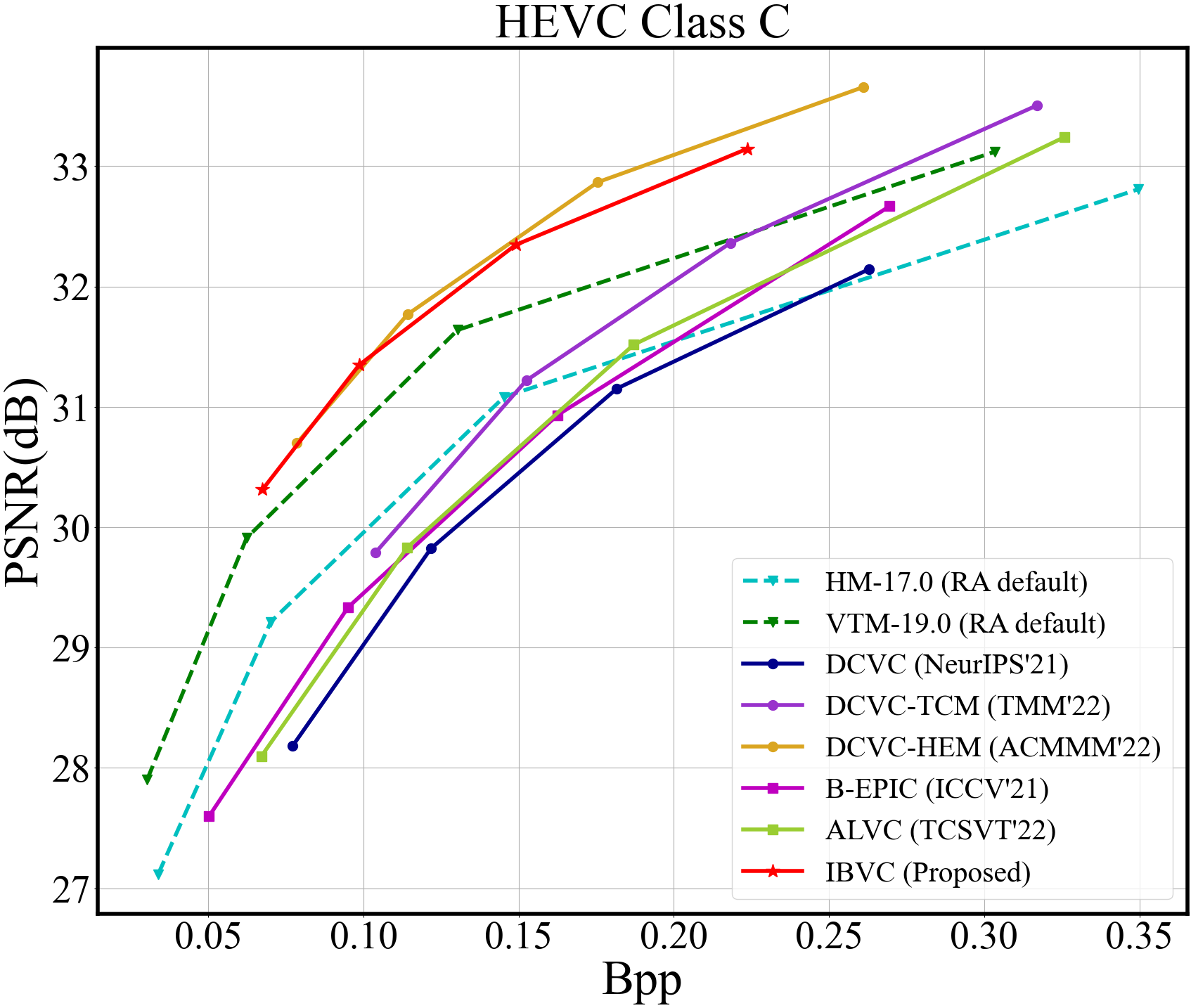}}
 \end{minipage}
 \vfill
 \begin{minipage}[b]{0.325\linewidth} 
   \centering
   \centerline{\includegraphics[width=\linewidth]{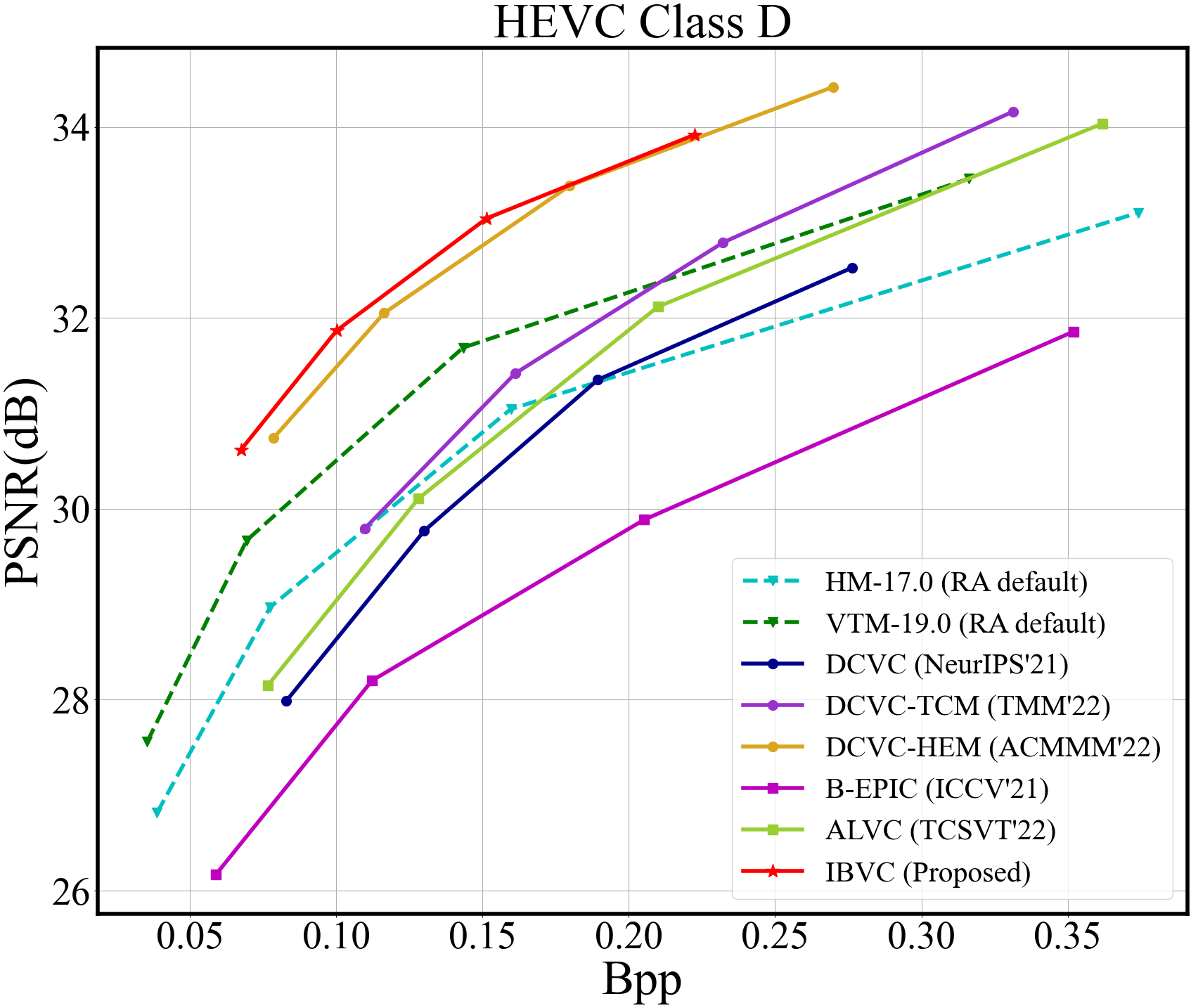}}
 \end{minipage}
 \hfill
 \begin{minipage}[b]{0.325\linewidth}
   \centering
   \centerline{\includegraphics[width=\linewidth]{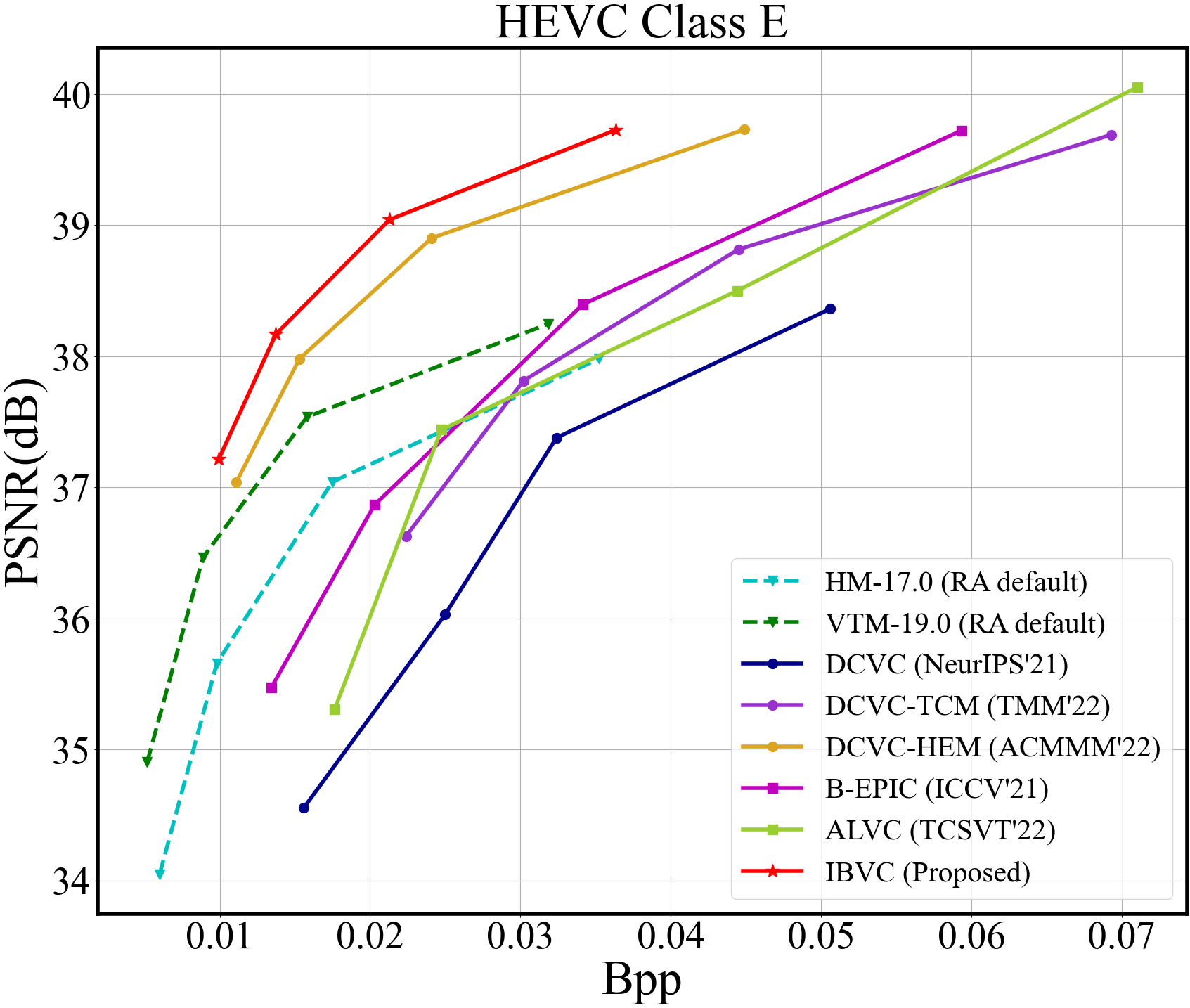}}
 \end{minipage}
 \hfill
 \begin{minipage}[b]{0.325\linewidth}
   \centering
   \centerline{\includegraphics[width=\linewidth]{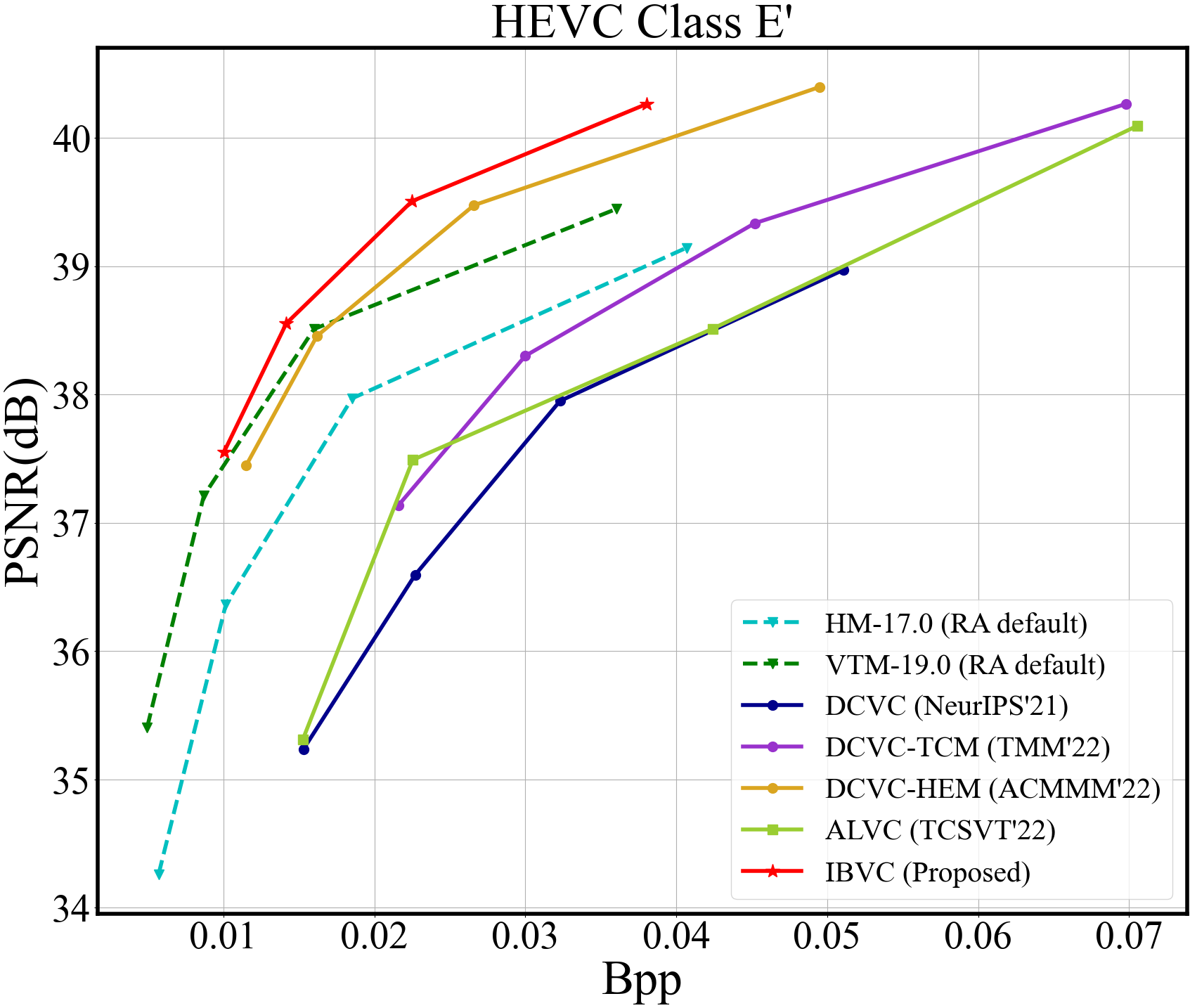}}
 \end{minipage}
   \caption{{Coding performance of IBVC evaluated on PSNR $\uparrow$.}}
  \label{fig5}
\end{figure*}

\begin{figure*}[!t]  
 \centering
 \begin{minipage}[b]{0.325\linewidth} 
   \centering
   \centerline{\includegraphics[width=\linewidth]{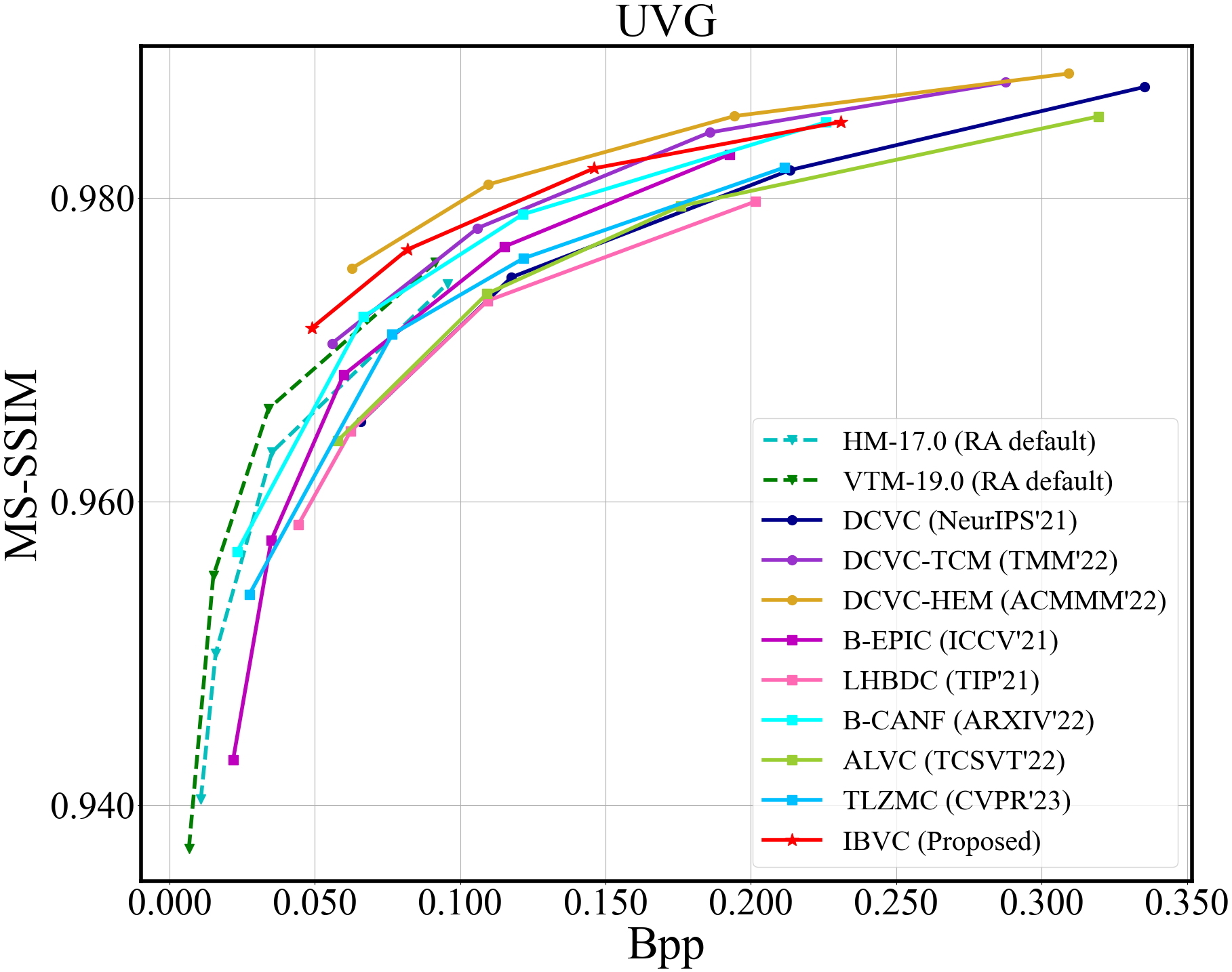}}
 \end{minipage}
 \hfill
 \begin{minipage}[b]{0.325\linewidth}
   \centering
   \centerline{\includegraphics[width=\linewidth]{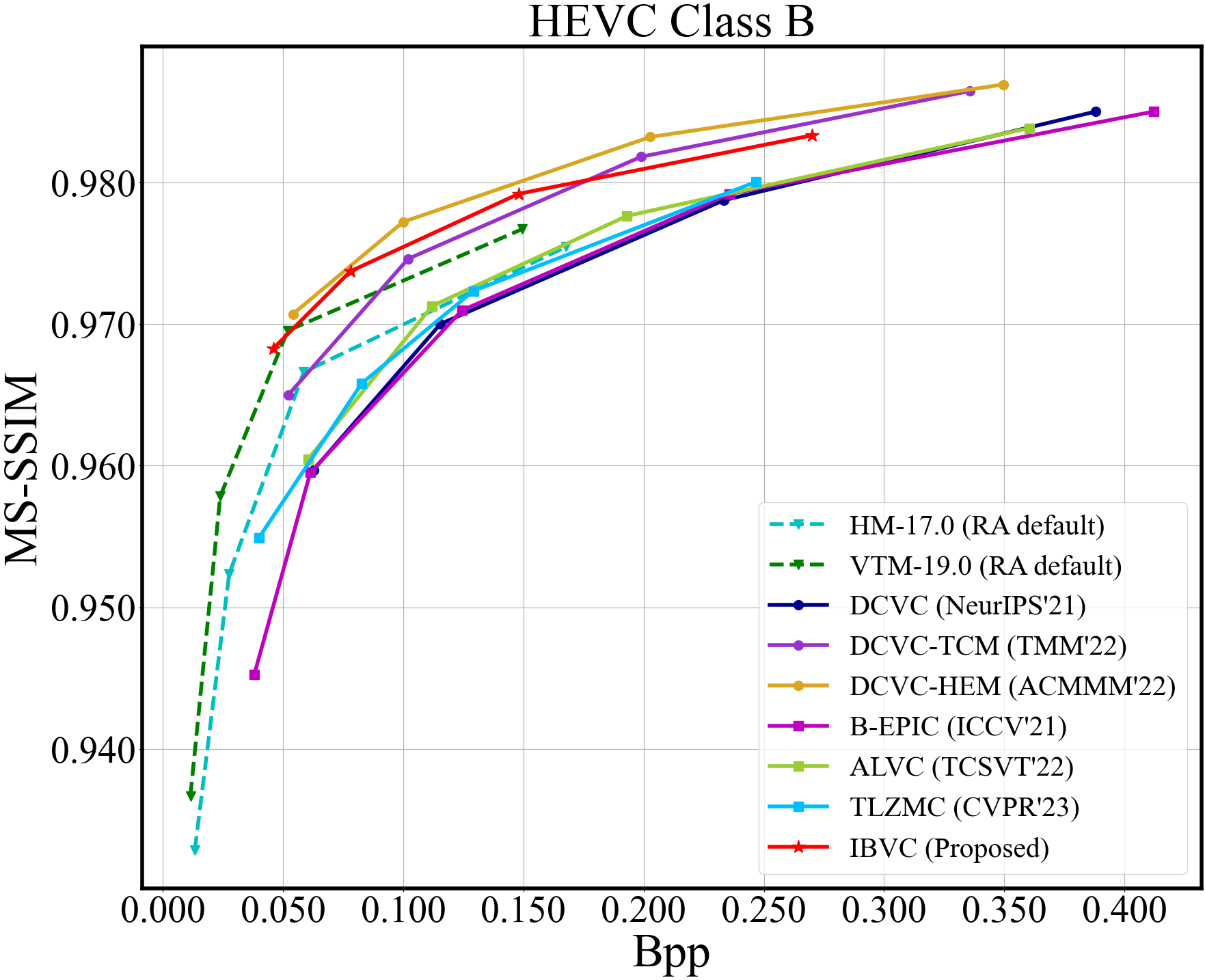}}
 \end{minipage}
 \hfill
 \begin{minipage}[b]{0.325\linewidth}
   \centering
   \centerline{\includegraphics[width=\linewidth]{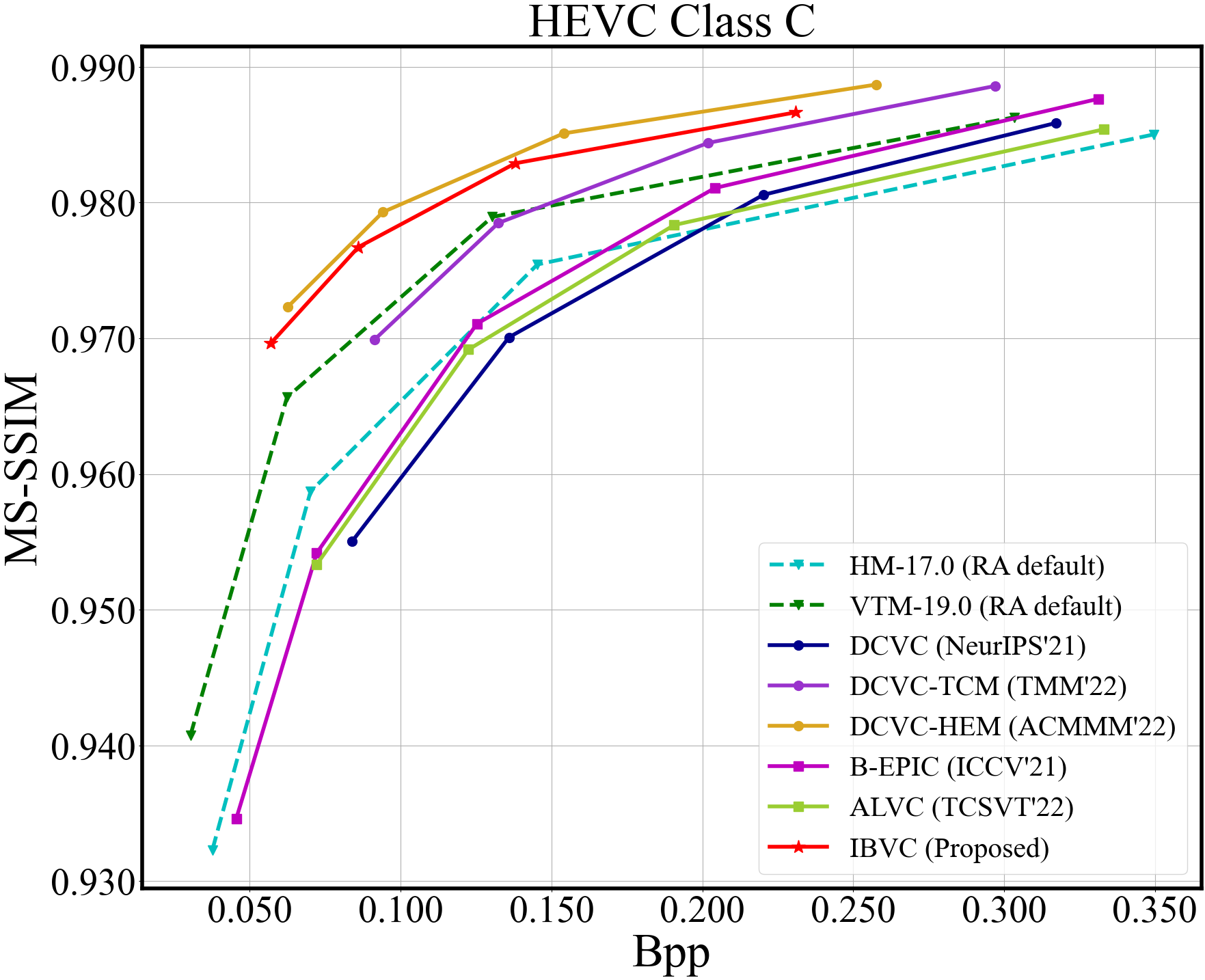}}
 \end{minipage}
 \vfill
 \begin{minipage}[b]{0.325\linewidth} 
   \centering
   \centerline{\includegraphics[width=\linewidth]{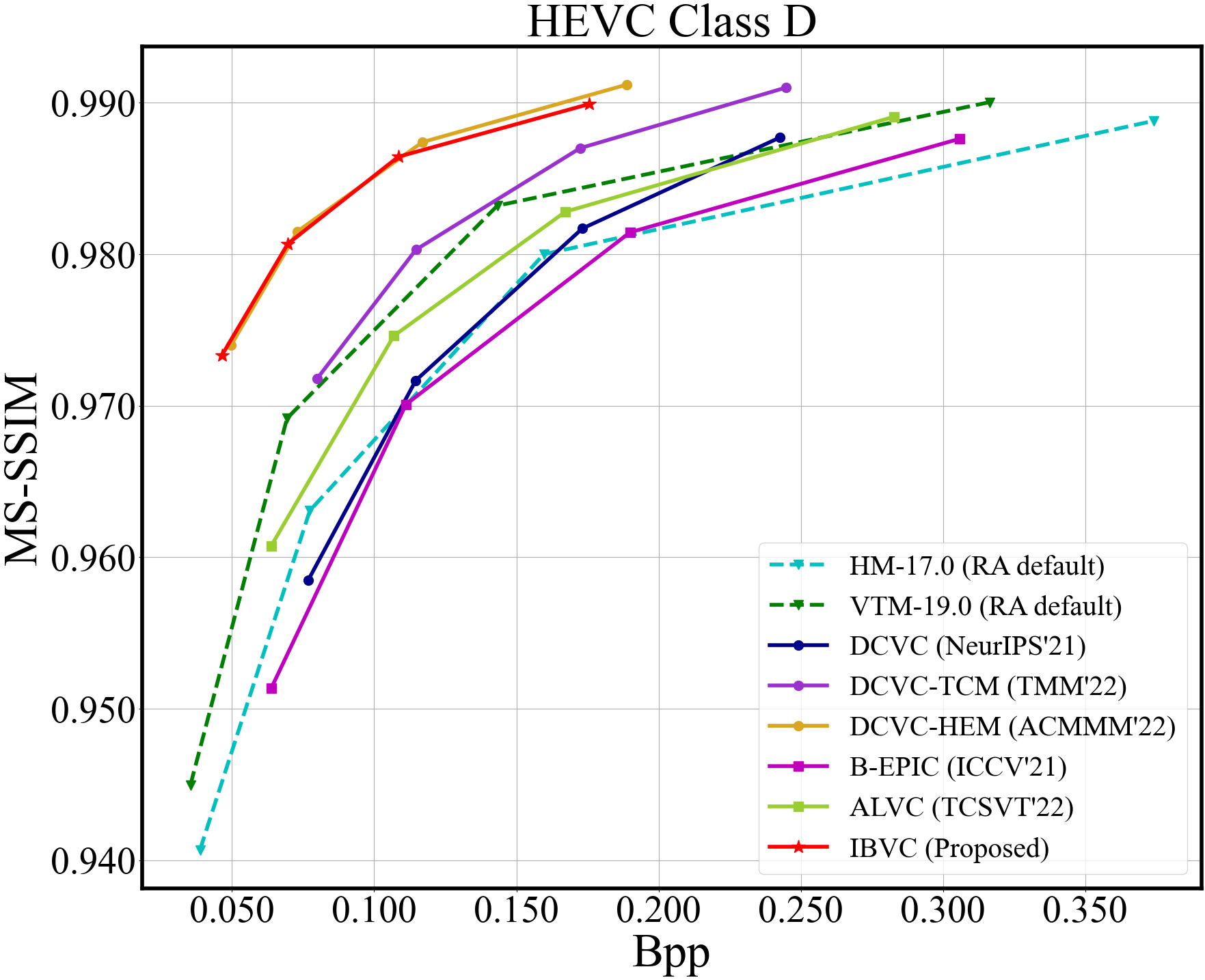}}
 \end{minipage}
 \hfill
 \begin{minipage}[b]{0.325\linewidth}
   \centering
   \centerline{\includegraphics[width=\linewidth]{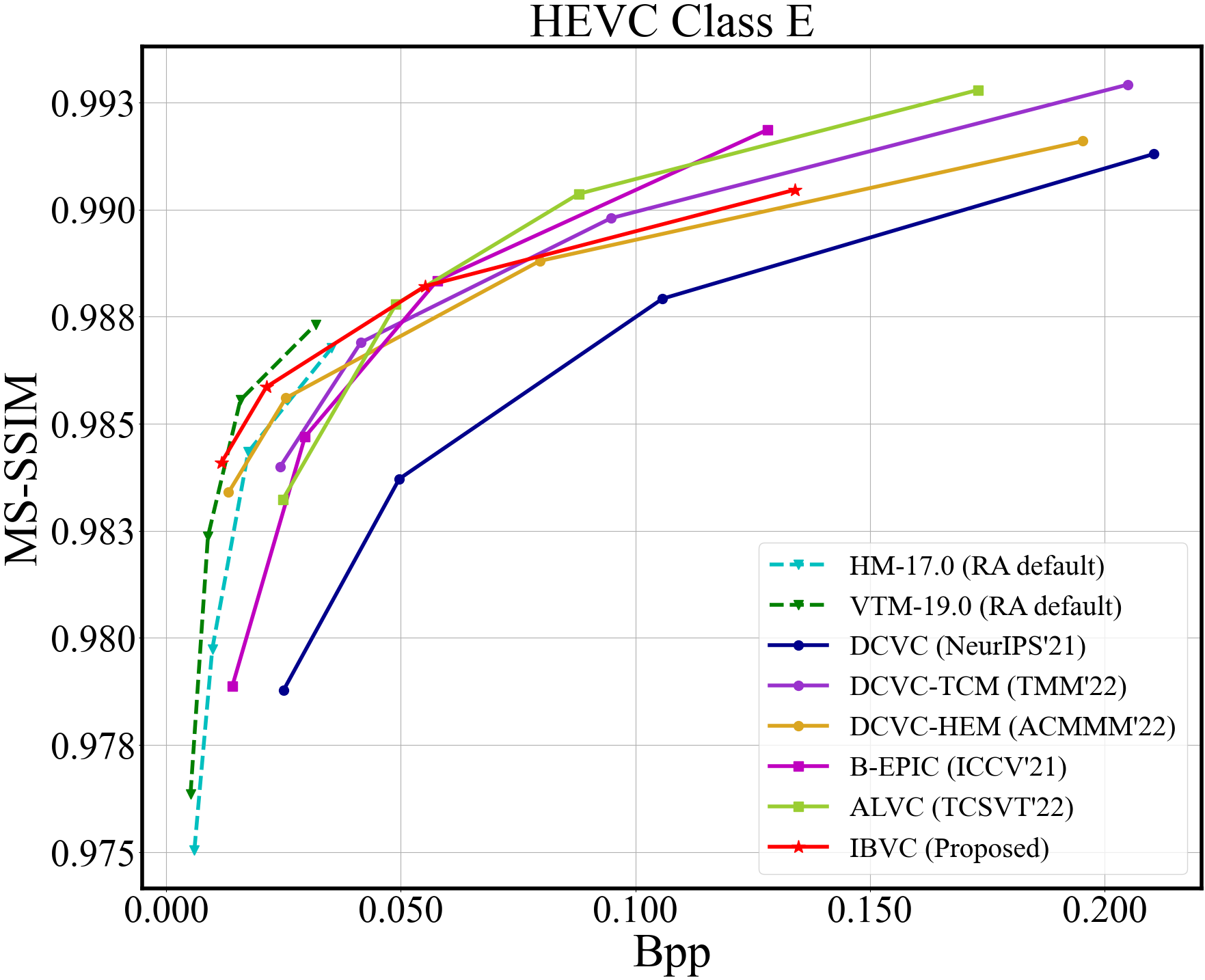}}
 \end{minipage}
 \hfill
 \begin{minipage}[b]{0.325\linewidth}
   \centering
   \centerline{\includegraphics[width=\linewidth]{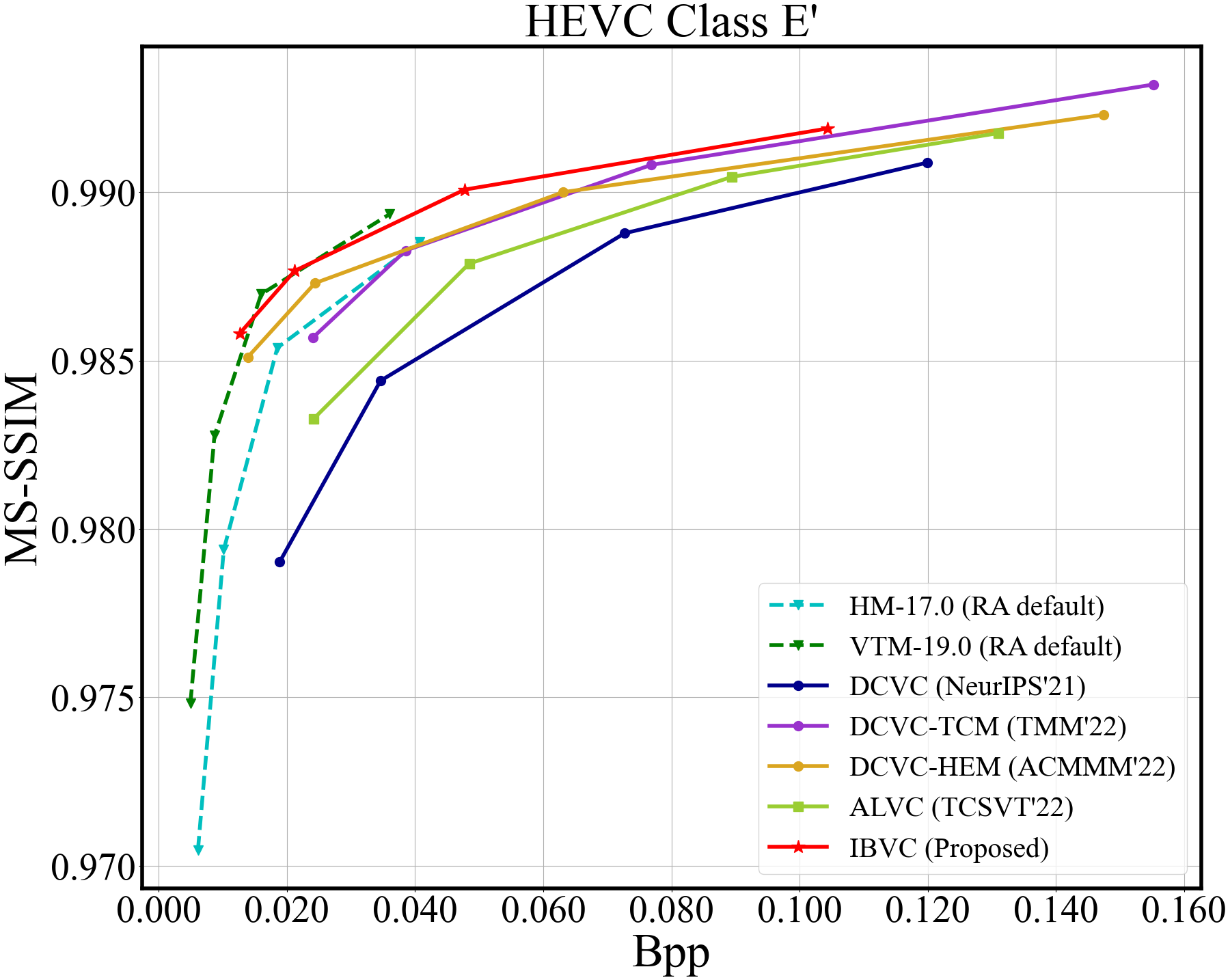}}
 \end{minipage}
   \caption{{Coding performance of IBVC evaluated on MS-SSIM $\uparrow$.}}
  \label{fig6}
\end{figure*}

\begin{figure*}[!t]  
 \centering
 \begin{minipage}[b]{0.325\linewidth} 
   \centering
   \centerline{\includegraphics[width=\linewidth]{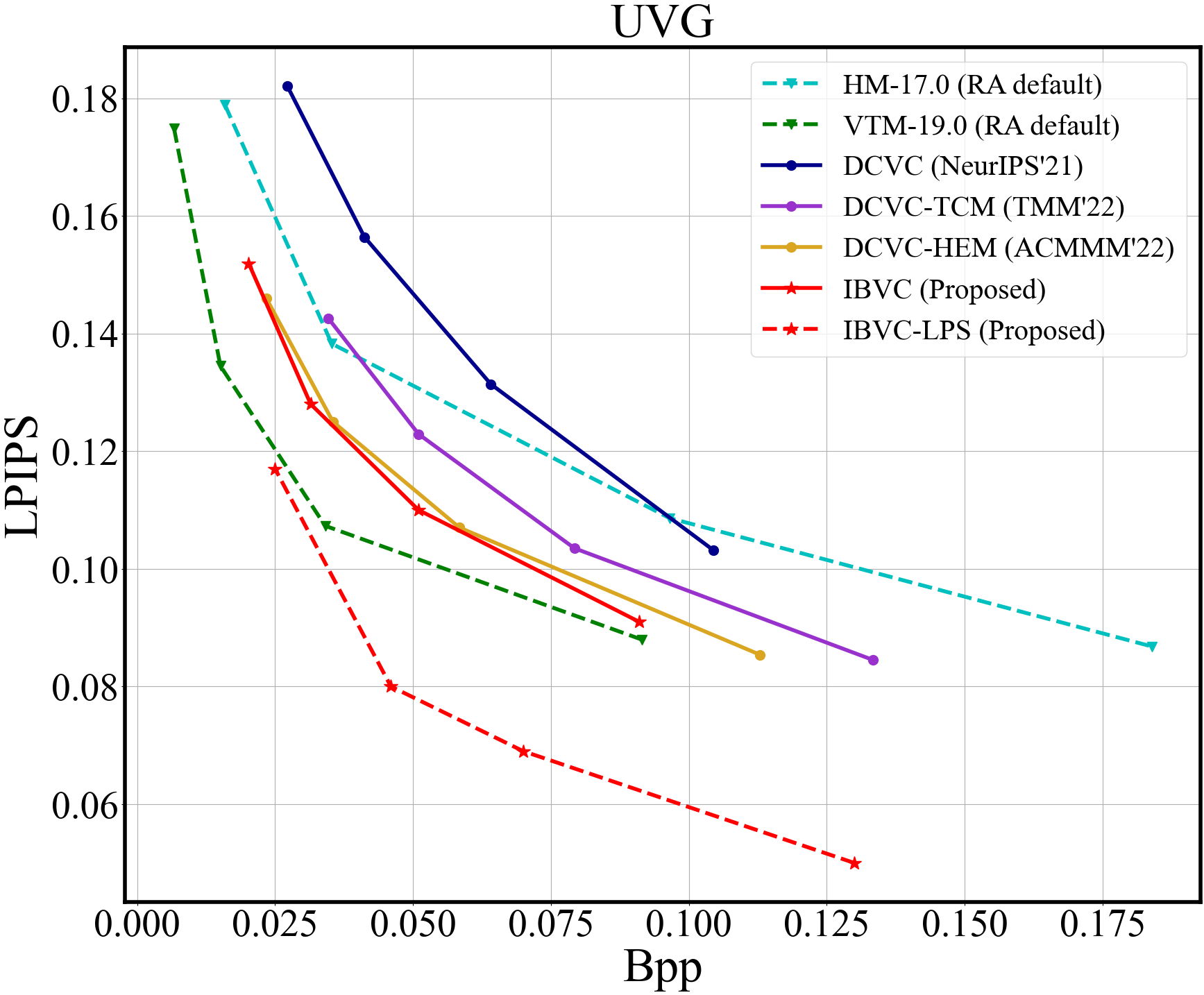}}
 \end{minipage}
 \hfill
 \begin{minipage}[b]{0.325\linewidth}
   \centering
   \centerline{\includegraphics[width=\linewidth]{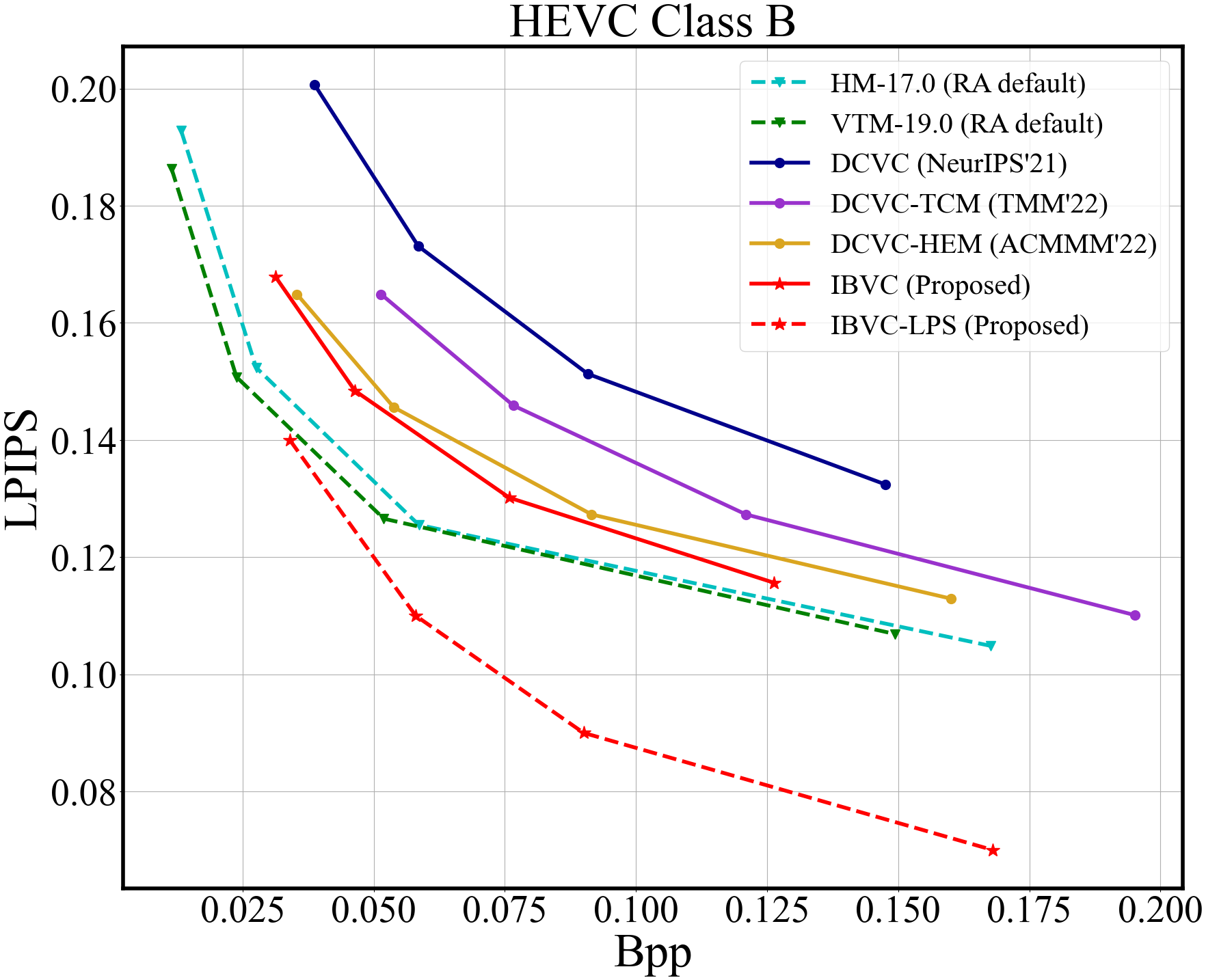}}
 \end{minipage}
 \hfill
 \begin{minipage}[b]{0.325\linewidth}
   \centering
   \centerline{\includegraphics[width=\linewidth]{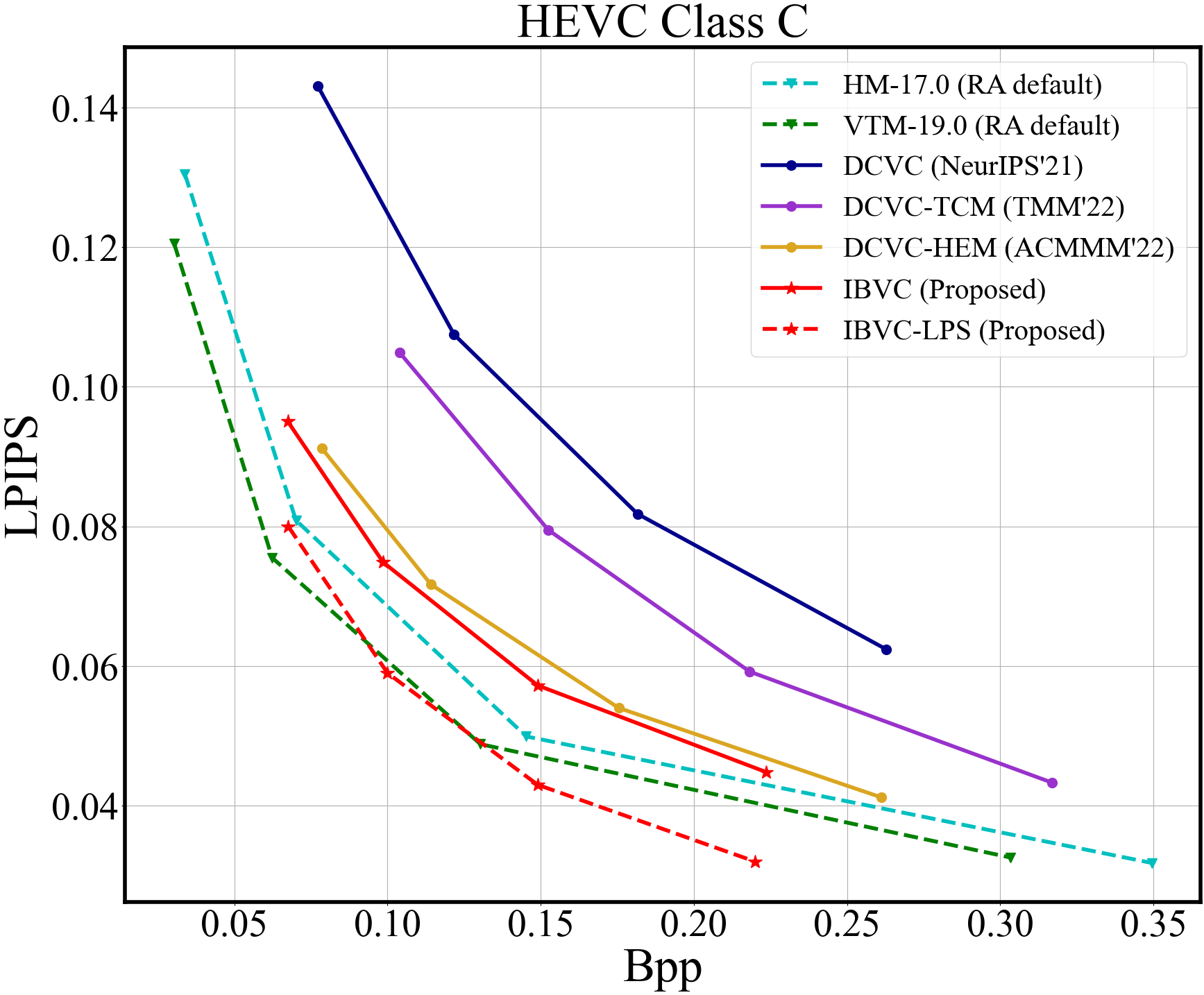}}
 \end{minipage}
 \vfill
 \begin{minipage}[b]{0.325\linewidth} 
   \centering
   \centerline{\includegraphics[width=\linewidth]{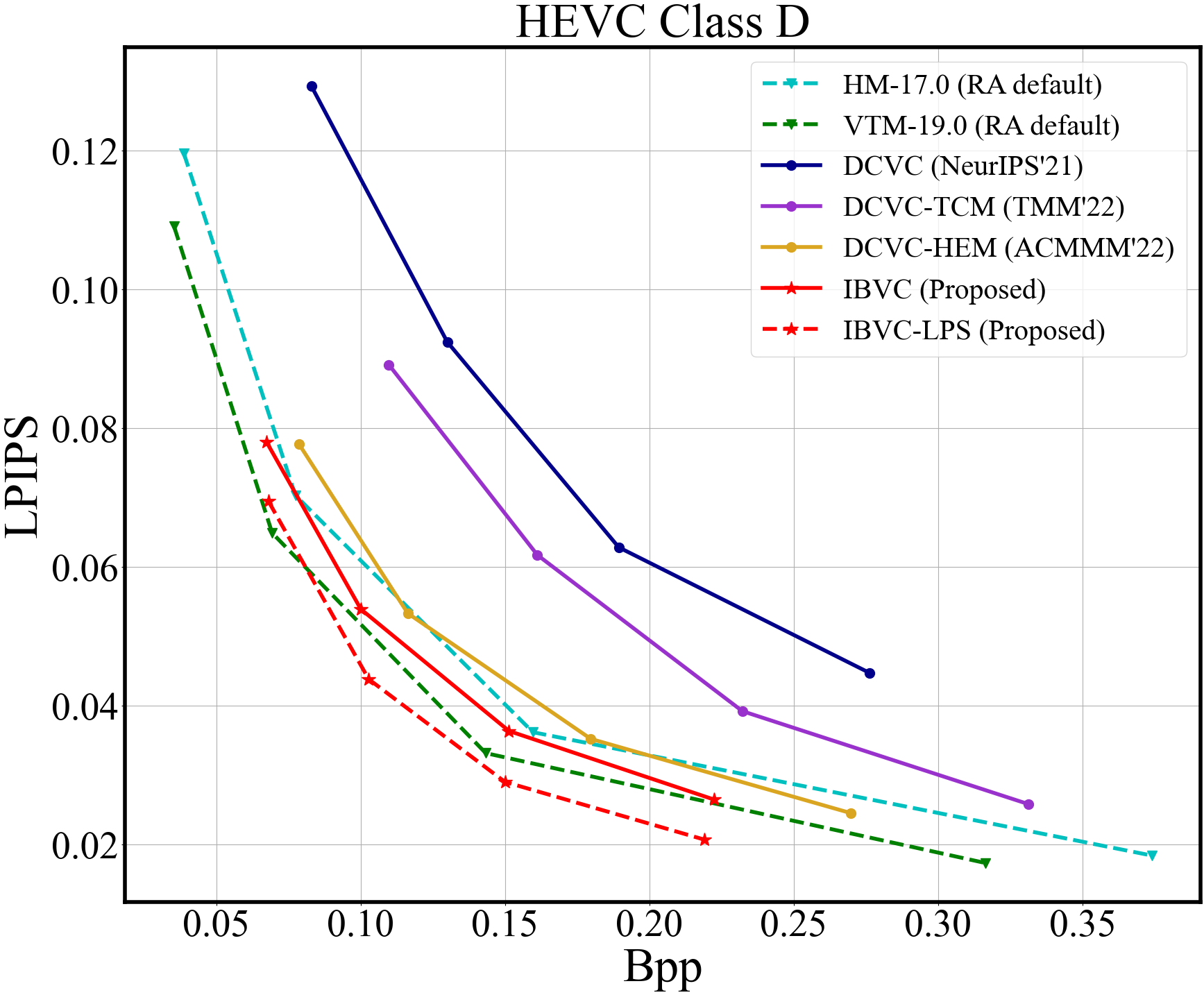}}
 \end{minipage}
 \hfill
 \begin{minipage}[b]{0.325\linewidth}
   \centering
   \centerline{\includegraphics[width=\linewidth]{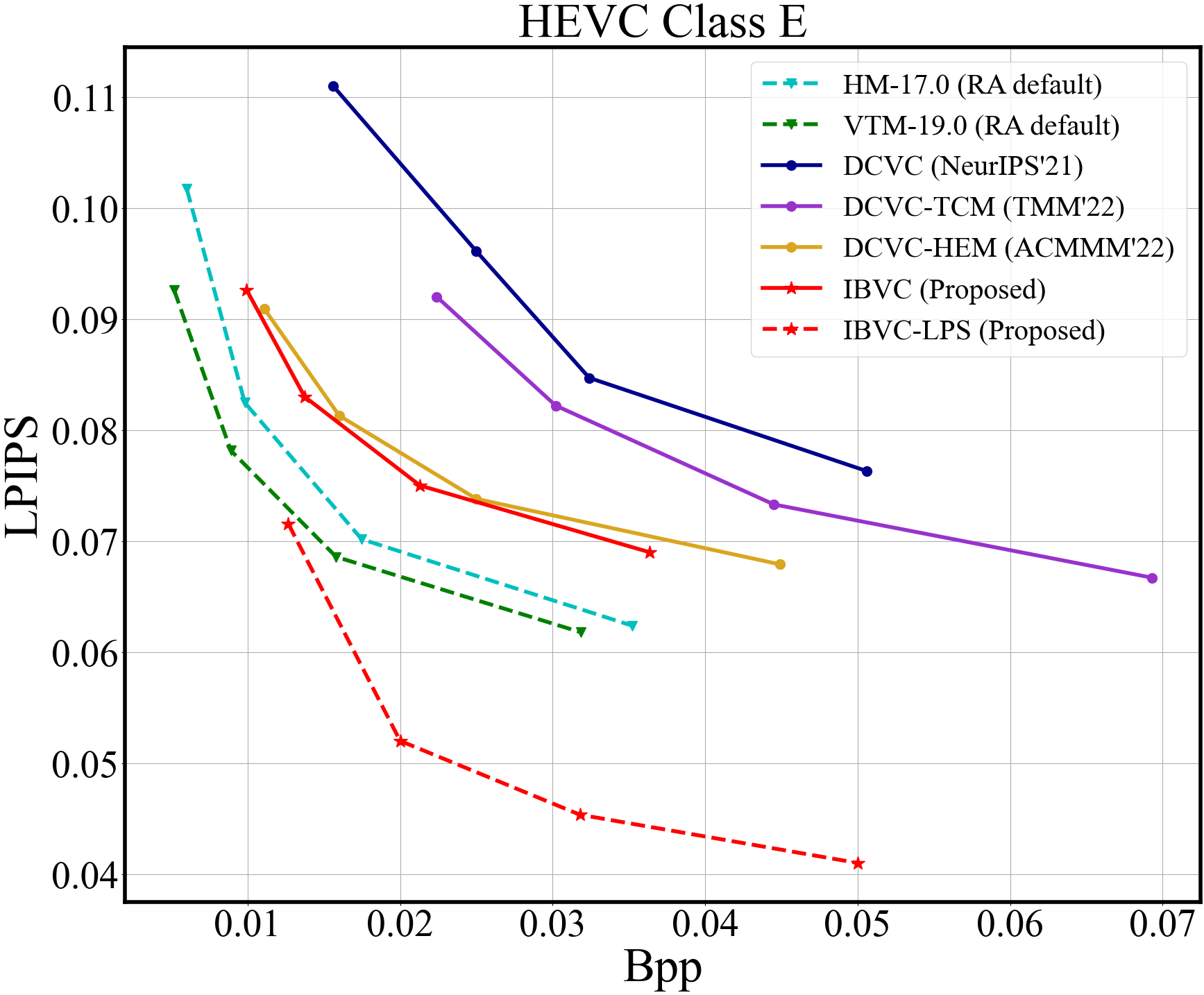}}
 \end{minipage}
 \hfill
 \begin{minipage}[b]{0.325\linewidth}
   \centering
   \centerline{\includegraphics[width=\linewidth]{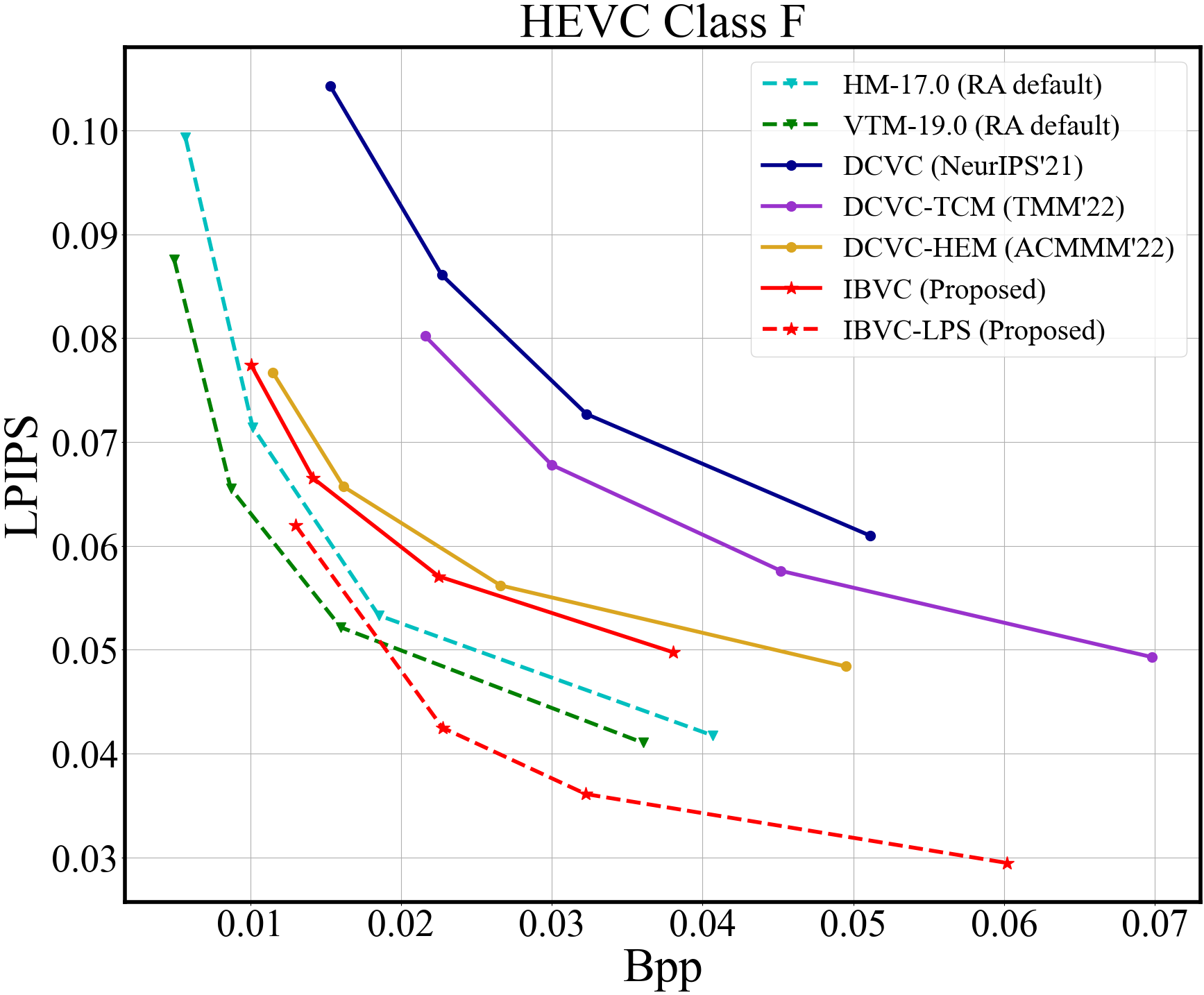}}
 \end{minipage}
   \caption{{Coding performance of IBVC evaluated on LPIPS $\downarrow$.}}
  \label{fig12}
\end{figure*}

\begin{table}
\setlength\tabcolsep{1.5pt}
\centering
\footnotesize
\caption{BD-rate (\%) $\downarrow$ comparison by PSNR. The anchor is VTM-19.0. \textbf{Bold} indicates the best results. \label{tab1}}
\begin{tabular}{ccccccc}
\hline
&VTM-19.0~\cite{vtm}  & HM-17.0~\cite{hm}  &B-EPIC~\cite{pourreza2021extending} & ALVC~\cite{yang2022Advancing} & TLZMC\cite{alexandre2023hierarchical} & IBVC \\
&(RA default)  & (RA default)  &(ICCV'21) & (TCSVT'22) & (CVPR'23) & (Proposed) \\
\hline
UVG & 0   &  38.31& 46.61   &  27.22 & 1.39 & \textbf{-48.41 }\\
Class B  & 0  & 46.77 & 102.73  & 103.00 & 88.88 & \textbf{-9.87} \\
Class C   & 0    & 44.57 &  70.54 & 64.31 & - & \textbf{-16.63}  \\
Class D  & 0    & 41.21 & 143.20 &  42.17 & -& \textbf{-36.24} \\
Class E  & 0    & 48.58  &  75.82  & 83.48  & -&  \textbf{-38.16} \\
Class E' & 0    &  54.43 &  -  &  131.54 & -&  \textbf{-15.78} \\
Average  & 0    & 45.65 &  87.78  & 75.28 & 45.14  & \textbf{-27.51}  \\
\hline
\end{tabular}
\end{table}

\begin{table}[!t]
\setlength\tabcolsep{1.5pt}
\centering
\footnotesize
\caption{BD-rate (\%) $\downarrow$ comparison by MS-SSIM. The anchor is VTM-19.0. \textbf{Bold} indicates the best results. \label{tab2}}
\begin{tabular}{ccccccc}
\hline
&VTM-19.0~\cite{vtm}  & HM-17.0~\cite{hm}  &B-EPIC~\cite{pourreza2021extending} & ALVC~\cite{yang2022Advancing} & TLZMC\cite{alexandre2023hierarchical} & IBVC \\
&(RA default)  & (RA default)  &(ICCV'21) & (TCSVT'22) & (CVPR'23) & (Proposed) \\
\hline
UVG & 0   &34.87&  84.06 & 71.03 & 53.62 &\textbf{-13.46}\\
Class B  & 0     & 40.14 &  4.42  & 85.14 & 0.60 &  \textbf{-5.07} \\
Class C   & 0   &40.37 & 48.25 &60.48 & -&\textbf{-20.62}  \\
Class D  & 0     &32.21 & 47.99 & 20.50 & -& \textbf{-40.72} \\
Class E  & \textbf{0}    &38.55 &  114.21 &107.58  & -&  54.39 \\
Class E' & \textbf{0}    & 49.35 &  - & 140.58 &- &   7.05 \\
Average  & 0     &39.25 &  59.78   &  80.88 & 27.11  &   \textbf{-30.07}    \\
\hline
\end{tabular}
\end{table}

\subsubsection{Quantitative Evaluation} 
It is noted from Figure~\ref{fig5} and Figure~\ref{fig6} that IBVC achieves state-of-the-art results on test datasets compared to other learned B-frame video compression methods in terms of PSNR and MS-SSIM R-D curves. We use Bpp (Bits per pixel) to measure the bits cost for one pixel in each frame. Especially, the PSNR models are also used for fair LPIPS evaluation. LPIPS is a perceptual metric for better objective visual performance measurement when using the same R-D loss. As shown in Figure~\ref{fig12}, a lower value corresponds to a better result, which indicates that IBVC achieves the best quantitative visual pleasing results on test datasets compared to other state-of-the-art learning-based coding methods. Additionally, we display the IBVC-LPS model trained with LPIPS distortion. The performance surpasses the traditional codecs HM-17.0~\cite{hm} and VTM-19.0~\cite{vtm}. {Besides, IBVC outperforms other learned P-frame video compression methods including DCVC-HEM~\cite{li2022hybrid}, especially in terms of the LPIPS metric. This improvement can be attributed to the composition of the GoP, which consists of B-frames using IBVC and I-frames and P-frames by DCVC-HEM. It is noted that the IBVC enhances the overall performance and the artifact reduction codec is helpful for human visual perception.} In Table~\ref{tab1} and Table~\ref{tab2}, we compare the BD-rate (\%) of VCEG-M33~\cite{VCEG-M33} for PSNR and MS-SSIM models. IBVC achieves an average bit-rate saving with the anchor of VTM-19.0, outperforming the other state-of-the-art B-frame coding methods.

\begin{figure}[!t]
\centering
\centerline{\includegraphics[width=\linewidth]{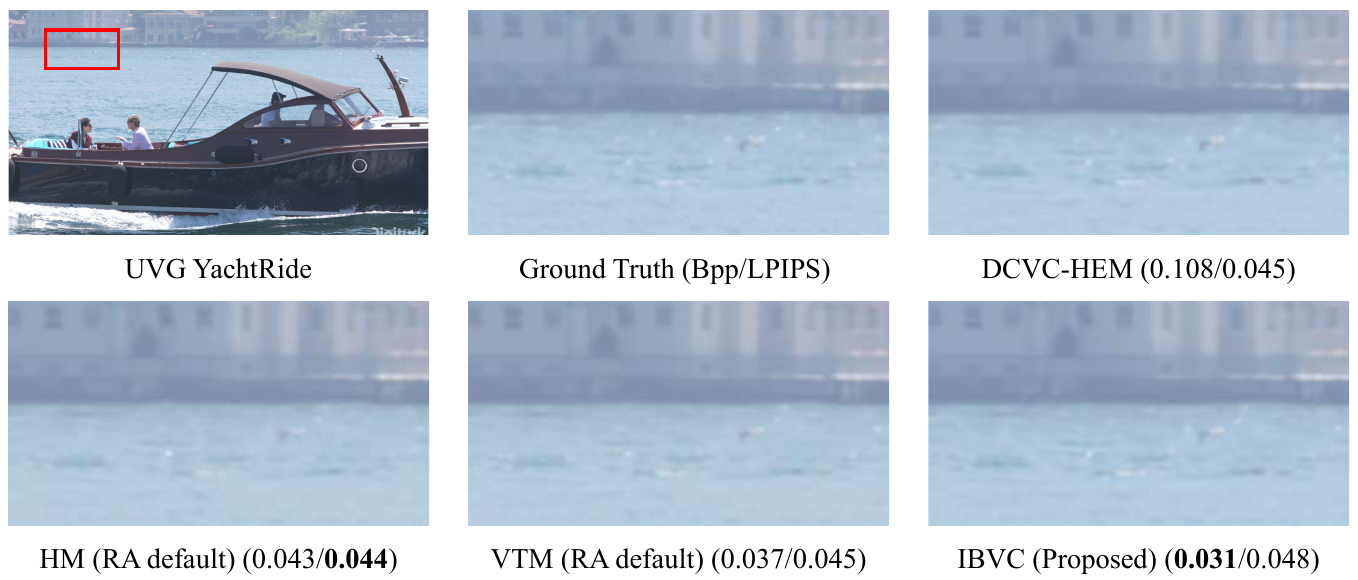}}
\caption{{Qualitative comparison on UVG YachtRide dataset.}}
\label{fig13}
\end{figure}

\begin{figure}[!t]
\centering
\centerline{\includegraphics[width=\linewidth]{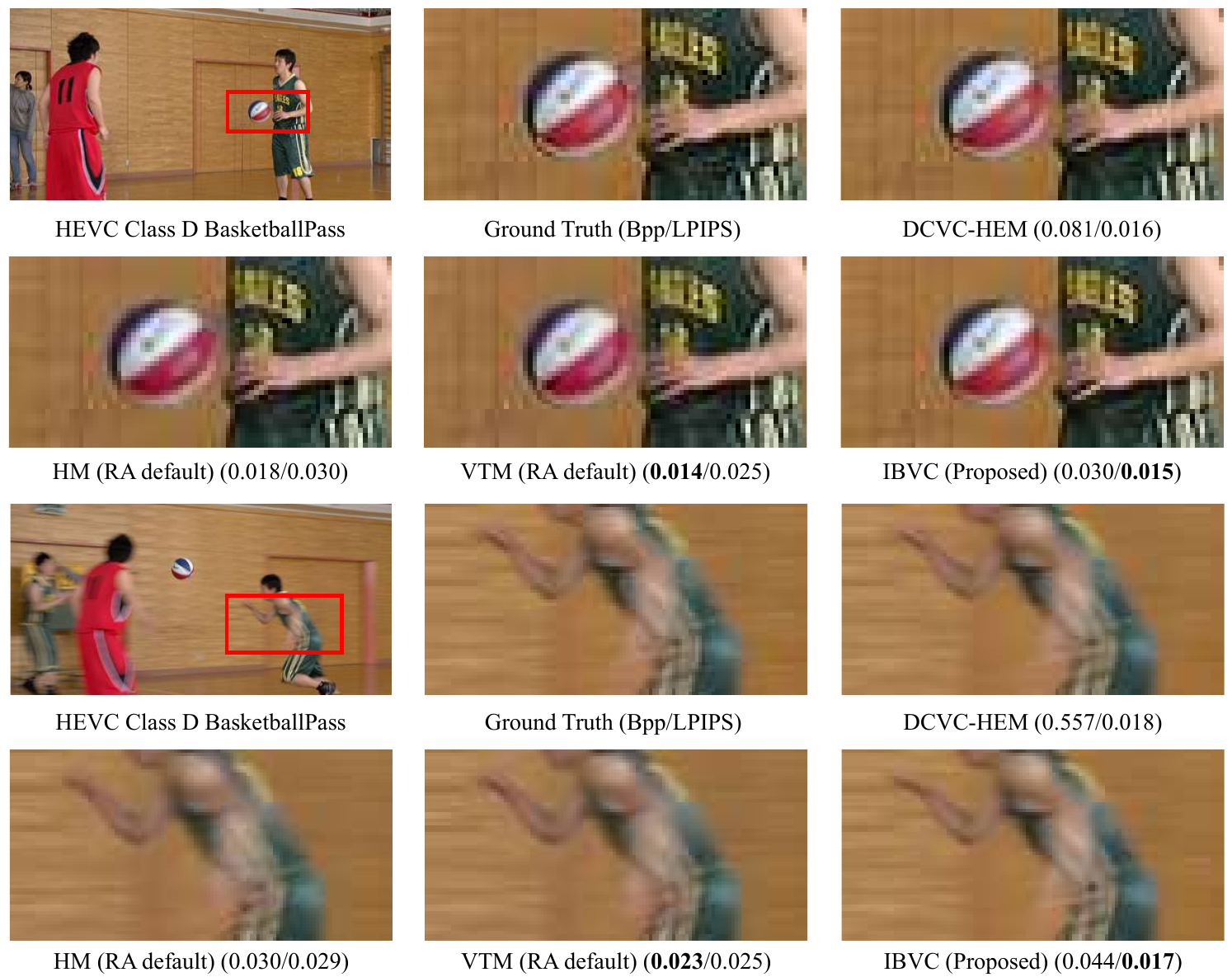}}
\caption{{Qualitative comparison on HEVC Class D BasketballPass dataset.}}
\label{fig14}
\end{figure}

\subsubsection{Qualitative Evaluation}
We provide a qualitative comparison of UVG and HEVC Class D datasets between our PSNR model and the latest state-of-the-art methods in Figure~\ref{fig13} and Figure~\ref{fig14}. To better demonstrate the performance of IBVC in artifact reduction, the LPIPS is used for perception evaluation in these examples. {As depicted in Figure~\ref{fig13}, IBVC achieves visually pleasing texture reconstruction compared to traditional codecs and has visually pleasing texture reconstruction compared to DCVC-HEM~\cite{li2022hybrid} with only one-third of the bit rates.} It can be observed that HM~\cite{hm} and VTM~\cite{vtm} lose some texture details of water ripples. Figure~\ref{fig14} displays two samples containing complex motion at low resolution. In contrast, our model achieves sharp boundaries and realistic textures without excessive smoothing and motion artifacts. Although it requires more bit rates than traditional codecs, the reconstruction performance exhibits fewer artifacts. {Compared to DCVC-HEM~\cite{li2022hybrid}, our method achieves lower LPIPS and better contextual reconstruction at a similar bit-rate.}

\begin{table}
\setlength\tabcolsep{3pt}
\centering
\footnotesize
\caption{Model parameters and BD-rate (\%) comparison by PSNR on UVG dataset. The anchor is VTM-19.0 with RA configuration. \textbf{Bold} indicates the best results.} \label{tab3}
\begin{tabular}{cccccc}
\hline
&LHBDC~\cite{yilmaz2021end} & FRHBVC~\cite{ccetin2022flexible} & B-CANF~\cite{chen2022b} & TLZMC~\cite{alexandre2023hierarchical} & IBVC \\
&(TIP'21) & (ICIP'22) & (TCSVT'23) & (CVPR'23) & (Proposed) \\
\hline
Parameters (M) $\downarrow$   & 23.5  & 35.0 & 24.0  & 39.9 & \textbf{11.6} \\
BD-rate (\%) $\downarrow$ & 17.13  & -16.49   &  -14.28 & 1.39 & \textbf{-48.41} \\
\hline
\end{tabular}
\end{table}

\begin{table}
\setlength\tabcolsep{1pt}
\centering
\footnotesize
\caption{{The comparison of coding time and BD-rate (\%) by PSNR on UVG dataset. The anchor is VTM-19.0 with RA configuration. \textbf{Bold} indicates the best results.} \label{tab4}}
\begin{tabular}{cccccc}
\hline
&DVC~\cite{lu2019dvc} & DCVC~\cite{li2021deep} &  DCVC-TCM~\cite{sheng2022temporal} &DCVC-HEM~\cite{li2022hybrid} & IBVC \\
&(CVPR'19) & (NeurIPS'21) & (TMM'22) & (ACMMM'22)  & (Proposed) \\
\hline
Coding time (ms) $\downarrow$   & 836  & 1564 & \textbf{533} & 643  & 832 \\
BD-rate (\%) $\downarrow$ & 120.64  & 47.19   & -15.41 &  -46.34 
 & \textbf{-48.41} \\
\hline
\end{tabular}
\end{table}

\subsubsection{Computational Complexity} 
We measure model parameters and BD-rate by PSNR of different methods. The evaluation results on 1080P videos from the UVG dataset are shown in Table~\ref{tab3}. Without redundant MEMC coding, IBVC achieves exceptional parameter saving over previous state-of-the-art approaches, with model parameters of 11.6M on B-frame coding, including the parameters of video frame interpolation method IFRNet-S~\cite{kong2022ifrnet}. {As shown in Table~\ref{tab4}, we measure the coding time and BD-rate by PSNR of different open-source methods on the same device. IBVC achieves a coding time of 832ms per frame, which includes artifact reduction coding and twice frame interpolation at the both encoder and decoder. Despite utilizing auto-regressive entropy models, IBVC is still comparable to the P-frame method DCVC-HEM~\cite{li2022hybrid}, which uses a checkerboard model, and surpasses the baseline method DCVC~\cite{li2021deep}.}

\begin{figure}[!t] 
 \centering
 \begin{minipage}[b]{0.325\linewidth} 
   \centering
   {\includegraphics[width=\linewidth]{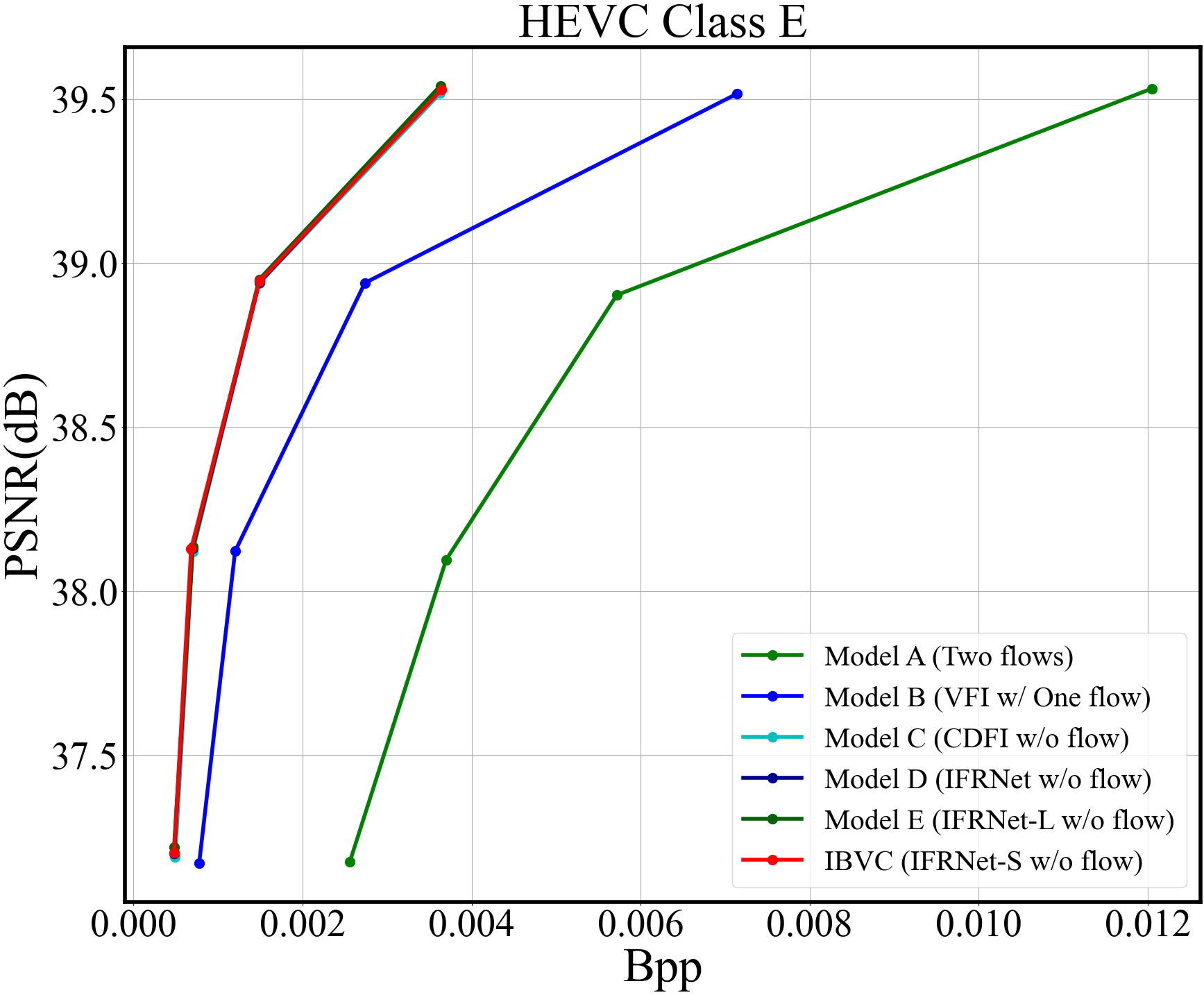}}
   {\footnotesize {(a) The comparison of MEMC strategies.}}
 \end{minipage}
 \hfill
 \begin{minipage}[b]{0.325\linewidth}
   \centering
   {\includegraphics[width=\linewidth]{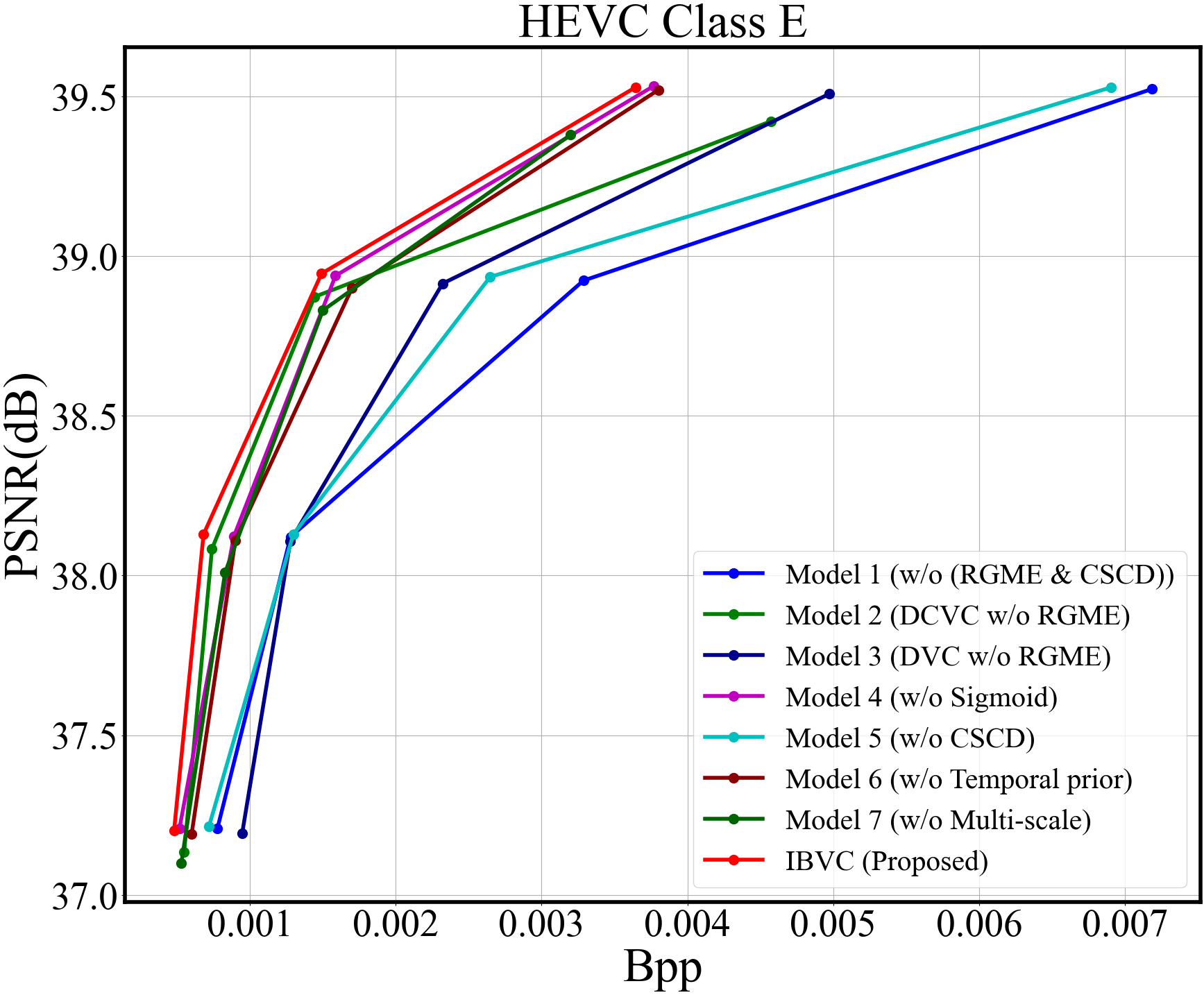}}
   {\footnotesize {(b) The effectiveness of the proposed codec.}}
 \end{minipage}
 \hfill
 \begin{minipage}[b]{0.325\linewidth}
   \centering
   {\includegraphics[width=\linewidth]{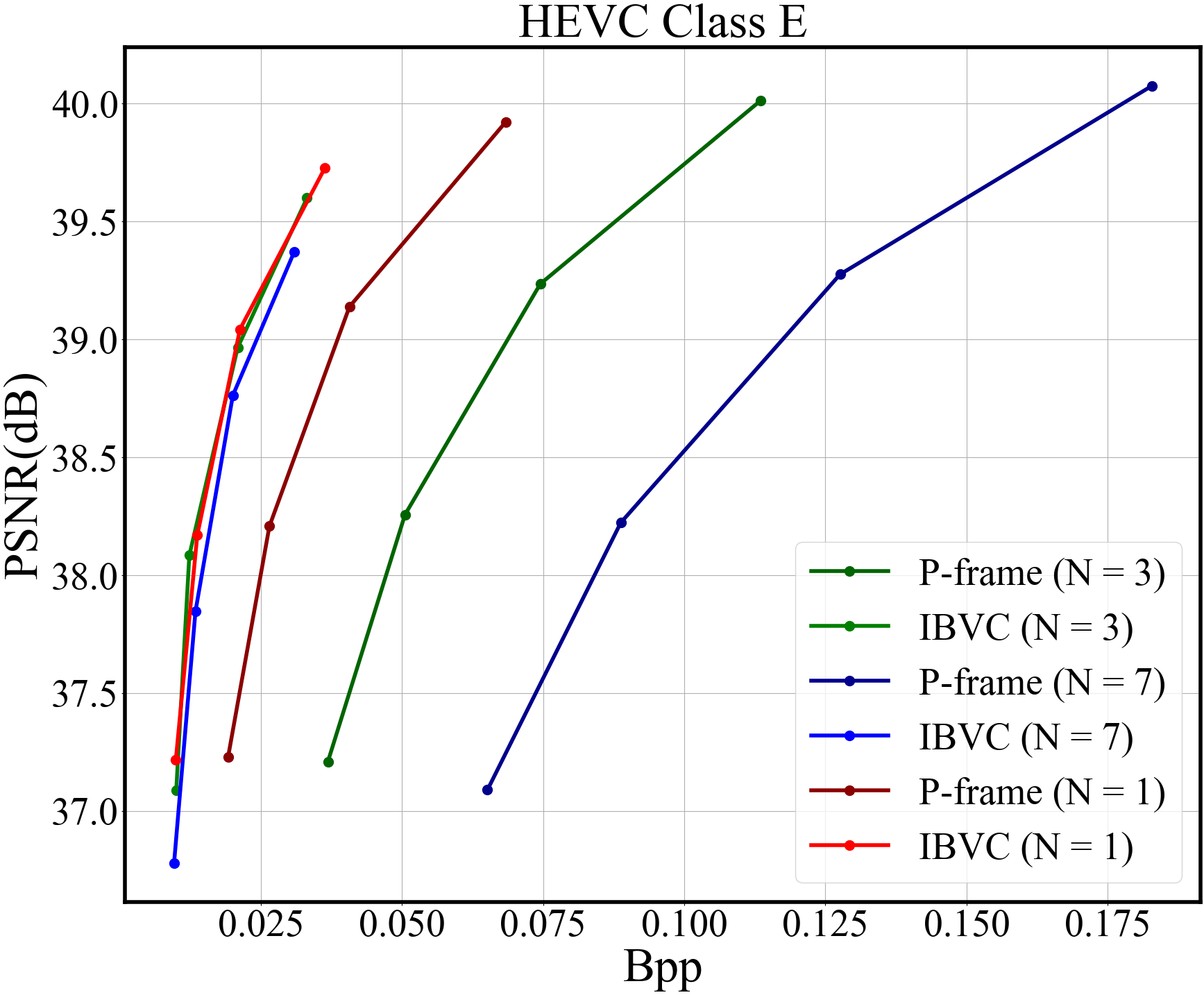}}
   {\footnotesize (c) The hierarchical B-frame quality.}
 \end{minipage}
   \caption{Ablation results of different MEMC strategies, proposed codec, and hierarchical B-frame quality. Only the performance of B-frame is shown in (a) and (b). Complete sequences are evaluated in (c). }
  \label{fig7}
\end{figure}

\subsection{Ablation Study}
\subsubsection{The Comparison of B-frame Coding Pipelines} 
One of our major contributions is the significant enhancement of computational efficiency by simplifying the B-frame coding pipeline. Concretely, to avoid duplicate MEMC coding operations, the optical-flow is not transmitted for alignment again after interpolation. To verify the effectiveness of the proposed pipeline, we compare it with the previous B-frame coding strategies, as shown in Figure~\ref{fig2}, using the same artifact reduction codec on the HEVC Class E dataset. As depicted in Figure~\ref{fig7}(a), Model A and Model B are designed to show that the proposed coding pipeline outperforms previous structures. Model A (Two flows) uses bi-directional flow transmission. Model B (VFI w/ One flow) includes frame interpolation with flow transmission. Moreover, IBVC is model-agnostic which can be easily combined with any video frame interpolation network. We utilize another VFI model CDFI~\cite{ding2021cdfi} to make ablation in Model C (CDFI w/o flow). {In addition, we perform frame interpolation using the higher-complexity IFRNet~\cite{kong2022ifrnet} and IFRNet-L~\cite{kong2022ifrnet} models as Model D and Model E.}

\begin{figure}[!t]
\centering{
\includegraphics[width=0.8\linewidth]{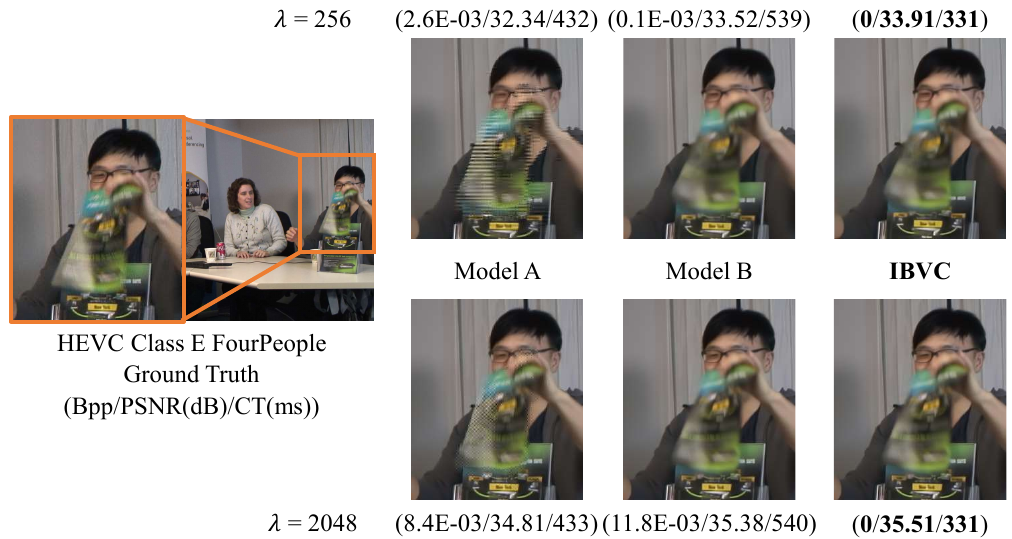}
}
\caption{Qualitative comparison of different MEMC strategies on HEVC Class E FourPeople dataset.}
\label{fig8}
\end{figure}

Our framework outperforms the bi-directional flow transmission framework (Model A) of Figure~\ref{fig2}(a) and the flow alignment framework (Model B) of Figure~\ref{fig2}(b). The qualitative evaluations and corresponding quantitative values of MEMC results before residual transmission are shown in Figure~\ref{fig8}, respectively. The results demonstrate that additional alignment can cause artifacts due to inaccurate flow estimation or quantification. Since not performing the additional MEMC coding compared with Model A and Model B, IBVC reduces the unnecessary coding time (CT) and new compression distortions. {Specifically, it can be observed that Model B exhibits checkerboard artifacts caused by quantization on the faces and books. In contrast, IBVC alleviates this phenomenon, indicating that the redundant motion codec causes noticeable artifacts from quantized optical flow in Equation~\ref{eq4}.} Besides, the replacement of the VFI method used deformable convolution in Model C does not affect coding performance, demonstrating the robustness of IBVC. {Model D and Model E, despite increasing the complexity of VFI, do not significantly improve the overall B-frame compression performance. It confirms the strong generalization capability of the codec in IBVC. As a result, IBVC utilizes the smaller IFRNet-S~\cite{kong2022ifrnet} for VFI.}

\subsubsection{The Effectiveness of Artifact Reduction Codec} 

Artifact reduction coding aims to capture differential spatio-temporal features between interpolated frames and input frames for compensating MEMC errors precisely. As described in Figure~\ref{fig7}(b), we compare the performance with different coding solutions on the HEVC Class E dataset. Model 1 (w/o (RGME \& CSCD)) employs the DCVC contextual codec to replace artifact reduction codec directly. Model 2 (DCVC w/o RGME) and Model 3 (DVC w/o RGME) respectively employ the DCVC~\cite{li2021deep} contextual encoder and DVC~\cite{lu2019dvc} residual encoder to replace artifact reduction encoder RGME. To further demonstrate the effectiveness of the mask in determining artifacts, we provide an ablation experiment Model 4 (w/o Sigmoid) without sigmoid function in RGME. Model 5 (w/o CSCD) replaces CSCD with the DCVC decoder. Model 6 (w/o Temporal prior) means not using prior conditions described in Equation~\ref{eq6}. {In Model 7 (w/o Multi-scale), we only extract multi-scale features $\hat{y}_t^1$ and $\hat{y}_t^l$ from $\hat{y}_t$ as described in Equation~\ref{eq11}.}

\begin{figure}[!t]
\centering
\centerline{\includegraphics[width=0.7\linewidth]{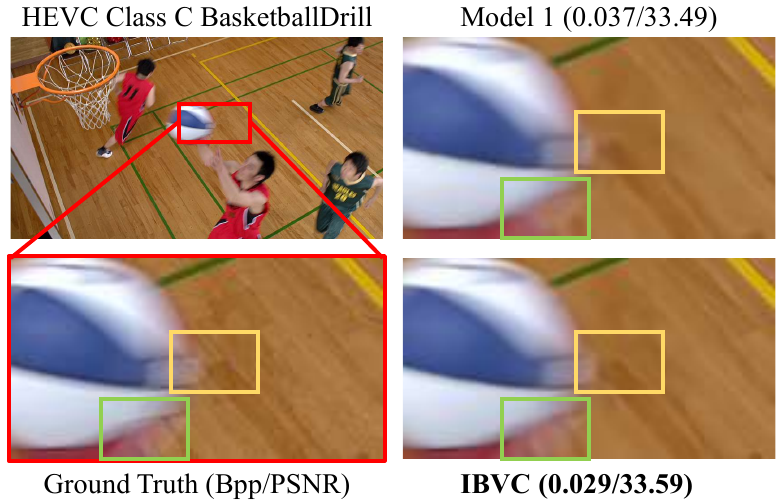}}
\caption{Visual example of artifact reduction from BasketballDrill in HEVC Class C dataset.}
\label{fig10}
\end{figure}

Model 1 shows great superiority of our artifact reduction codec in B-frame reconstruction after VFI. It is noteworthy that we selectively learn informative features reducing the bit-rate consumption from Model 2 and Model 3. Model 4 demonstrates that the sigmoid function can aggregate artifact information to reduce bit rates. Model 5 verifies that the spatio-temporal decoder has an influence on exquisite frame reconstruction. Model 6 shows that the inter-frame spatio-temporal information is useful for minimizing MEMC errors. {Model 7 shows that extracting complete multi-scale features from $\hat{y}_t$ improves the reconstruction performance.} The example in Figure~\ref{fig10} shows the effective reduction of artifacts. From these comparisons, our artifact reduction codec can achieve clearer structures and fewer distortions in marked locations without the extra bit-rate cost.

\subsubsection{Hierarchical B-frame Quality} 
To investigate the influence of the reference units on hierarchical B-frame quality, we vary the number of consecutive B-frames ($N$) between two reference frames on the HEVC Class E dataset. In the default experiment, we set the $N = 1$ with a reference unit size of 3, which yields the best performance and improvement on the reference P-frame sequences. The reference P-frame sequences are compressed by DCVC-HEM~\cite{li2022hybrid}. To ensure temporal consistency in consecutive B-frames, the transmitted B-frames are treated as the reference frames to generate new hierarchical B-frames through a recursive relation. Thus, the number of B-frames must follow the rule:
\begin{equation}
 N = \sum_0^i 2^i ~~~ i = 0,1,2, ...
\label{eq14}
\end{equation}
where $i$ means the hierarchical level. We also test with three and seven consecutive B-frames ($N = 3,7$) in reference units. For instance, P-frame ($N = 3$) shows the performance of only the reference P-frame sequences, where reference units have three consecutive B-frames. It is observed that the B-frames boost the performance compared with the reference P-frame sequences from Figure~\ref{fig7}(c). However, with an increase in the number of B-frames, the distance between two reference frames becomes longer. It is robust to use the $N = 1$ with a reference unit size of 3. Summarily, IBVC can improve P-frame performance in the hierarchical structure of reference units and is not limited by different GoP structures.

\section{Conclusion}

We have proposed a simple yet efficient interpolation-driven B-frame video compression (IBVC) algorithm with excellent efficiency and performance. IBVC is based on the analysis of previous methods that the repetitive alignment in MEMC cannot improve the compression ratio and produce temporal redundancy with additional artifacts. We directly perform artifact reduction coding after frame interpolation, drastically reducing computational costs with superior performance. In addition, an artifact reduction encoder-decoder network is designed to adaptively select the most useful contexts and construct a transmitted representation by prior intra/inter-frame dependencies. Experimental results demonstrate that IBVC significantly reduces both computational costs and bit-rate requirements with excellent reconstruction performance, compared with the relevant state-of-the-art B-frame video compression methods. {IBVC relies exclusively on the artifact reduction codec for transmission, which requires a stable auto-regressive entropy model. Consequently, IBVC exhibits slightly lower efficiency compared to P-frame compression employing a checkerboard entropy model. As a result, future research will involve exploring a bi-directional entropy coding strategy that draws inspiration from bi-directional inter-frame prediction to improve the estimation of the probability distribution for intermediate frames.}

\section{Acknowledgements}
This work is supported in part by the National Key Research and Development Program of China under Grant 2022ZD0118001; and in part by the National Natural Science Foundation of China under Grant 62372036, Grant 61972028, Grant 62332017, and Grant 62120106009.


\bibliographystyle{elsarticle-num} 
\bibliography{ref}

\begin{thebibliography}{10}
\expandafter\ifx\csname url\endcsname\relax
  \def\url#1{\texttt{#1}}\fi
\expandafter\ifx\csname urlprefix\endcsname\relax\def\urlprefix{URL }\fi
\expandafter\ifx\csname href\endcsname\relax
  \def\href#1#2{#2} \def\path#1{#1}\fi

\bibitem{wang2023Versatile}
Y.~Wang, X.~Bai, Versatile recurrent neural network for wide types of video restoration, {Pattern Recognit.} 138 (2023) 109360.

\bibitem{chen2024High}
R.~Chen, Y.~Mu, Y.~Zhang, High-order relational generative adversarial network for video super-resolution, {Pattern Recognit.} 146 (2024) 110059.

\bibitem{PATIL2022Dual}
P.~W. Patil, S.~Gupta, S.~Rana, S.~Venkatesh, Dual-frame spatio-temporal feature modulation for video enhancement, {Pattern Recognit.} 130 (2022) 108822.

\bibitem{sheng2024vnvc}
X.~Sheng, L.~Li, D.~Liu, H.~Li, {VNVC}: A versatile neural video coding framework for efficient human-machine vision, {IEEE Trans. on Pattern Anal. and Mach. Intell.} (2024).

\bibitem{Liu2021Mutual}
X.~Liu, L.~Jin, X.~Han, J.~You, Mutual information regularized identity-aware facial expression recognition in compressed video, {Pattern Recognit.} 119 (2021) 108105.

\bibitem{Qiao2021Deep}
S.~Qiao, R.~Wang, S.~Shan, X.~Chen, Deep video code for efficient face video retrieval, {Pattern Recognit.} 113 (2021) 107754.

\bibitem{Shohei2023Deep}
S.~Uchigasaki, T.~Miyazaki, S.~Omachi, Deep image compression using scene text quality assessment, {Pattern Recognit.} 142 (2023) 109696.

\bibitem{lu2019dvc}
G.~Lu, W.~Ouyang, D.~Xu, X.~Zhang, C.~Cai, Z.~Gao, {DVC}: An end-to-end deep video compression framework, in: {Proceedings of the IEEE/CVF Conference on Computer Vision and Pattern Recognition}, 2019, pp. 11006--11015.

\bibitem{li2021deep}
J.~Li, B.~Li, Y.~Lu, Deep contextual video compression, in: {Advances in Neural Information Processing Systems}, Vol.~34, 2021, pp. 18114--18125.

\bibitem{bross2021overview}
B.~Bross, Y.-K. Wang, Y.~Ye, S.~Liu, J.~Chen, G.~J. Sullivan, J.-R. Ohm, Overview of the versatile video coding ({VVC}) standard and its applications, {IEEE Trans. on Circuits and Syst. for Video Technol.} 31~(10) (2021) 3736--3764.

\bibitem{yang2022Advancing}
R.~Yang, R.~Timofte, L.~Van~Gool, Advancing learned video compression with in-loop frame prediction, {IEEE Trans. on Circuits and Syst. for Video Technol.} 33~(5) (2023) 2410--2423.

\bibitem{yang2020Learning}
R.~Yang, F.~Mentzer, L.~Van~Gool, R.~Timofte, Learning for video compression with hierarchical quality and recurrent enhancement, in: {Proceedings of the IEEE/CVF Conference on Computer Vision and Pattern Recognition}, 2020, pp. 6628--6637.

\bibitem{pourreza2021extending}
R.~Pourreza, T.~Cohen, Extending neural {P}-frame codecs for {B}-frame coding, in: {Proceedings of the IEEE/CVF International Conference on Computer Vision}, 2021, pp. 6680--6689.

\bibitem{balle2017end}
J.~Ball{\'e}, V.~Laparra, E.~P. Simoncelli, End-to-end optimized image compression, in: {Proceedings of the International Conference on Learning Representations}, 2017, pp. 1--27.

\bibitem{cheng2020learned}
Z.~Cheng, H.~Sun, M.~Takeuchi, J.~Katto, Learned image compression with discretized {Gaussian} mixture likelihoods and attention modules, in: {Proceedings of the IEEE/CVF Conference on Computer Vision and Pattern Recognition}, 2020, pp. 7939--7948.

\bibitem{sheng2022temporal}
X.~Sheng, J.~Li, B.~Li, L.~Li, D.~Liu, Y.~Lu, Temporal context mining for learned video compression, {IEEE Trans. on Multimed.} (2022) 1--12.

\bibitem{guo2023learning}
Z.~Guo, R.~Feng, Z.~Zhang, X.~Jin, Z.~Chen, Learning cross-scale weighted prediction for efficient neural video compression, {IEEE Trans. on Image Process.} 32 (2023) 3567--3579.

\bibitem{li2022hybrid}
J.~Li, B.~Li, Y.~Lu, Hybrid spatial-temporal entropy modelling for neural video compression, in: {Proceedings of the ACM International Conference on Multimedia}, 2022, pp. 1503--1511.

\bibitem{djelouah2019neural}
A.~Djelouah, J.~Campos, S.~Schaub-Meyer, C.~Schroers, Neural inter-frame compression for video coding, in: {Proceedings of the IEEE/CVF International Conference on Computer Vision}, 2019, pp. 6421--6429.

\bibitem{ccetin2022flexible}
E.~{\c{C}}etin, M.~A. Y{\i}lmaz, A.~M. Tekalp, Flexible-rate learned hierarchical bi-directional video compression with motion refinement and frame-level bit allocation, in: {Proceedings of the IEEE International Conference on Image Processing}, 2022, pp. 1206--1210.

\bibitem{chen2022b}
M.-J. Chen, Y.-H. Chen, W.-H. Peng, {B-CANF}: Adaptive {B-frame} coding with conditional augmented normalizing flows, {IEEE Trans. on Circuits and Syst. for Video Technol.} (2023) 1--14.

\bibitem{kong2022ifrnet}
L.~Kong, B.~Jiang, D.~Luo, W.~Chu, X.~Huang, Y.~Tai, C.~Wang, J.~Yang, {IFRNet}: Intermediate feature refine network for efficient frame interpolation, in: {Proceedings of the IEEE/CVF Conference on Computer Vision and Pattern Recognition}, 2022, pp. 1969--1978.

\bibitem{ding2021cdfi}
T.~Ding, L.~Liang, Z.~Zhu, I.~Zharkov, {CDFI}: Compression-driven network design for frame interpolation, in: {Proceedings of the IEEE/CVF Conference on Computer Vision and Pattern Recognition}, 2021, pp. 8001--8011.

\bibitem{jiang2018super}
H.~Jiang, D.~Sun, V.~Jampani, M.-H. Yang, E.~Learned-Miller, J.~Kautz, {Super SloMo}: High quality estimation of multiple intermediate frames for video interpolation, in: {Proceedings of the IEEE/CVF Conference on Computer Vision and Pattern Recognition}, 2018, pp. 9000--9008.

\bibitem{wu2018video}
C.-Y. Wu, N.~Singhal, P.~Krahenbuhl, Video compression through image interpolation, in: {Proceedings of the European Conference on Computer Vision}, 2018, pp. 416--431.

\bibitem{jia2022neighbor}
Z.~Jia, Y.~Lu, H.~Li, Neighbor correspondence matching for flow-based video frame synthesis, in: {Proceedings of the ACM International Conference on Multimedia}, 2022, pp. 5389--5397.

\bibitem{alexandre2023hierarchical}
D.~Alexandre, H.-M. Hang, W.-H. Peng, Hierarchical b-frame video coding using two-layer canf without motion coding, in: {Proceedings of the IEEE/CVF Conference on Computer Vision and Pattern Recognition}, 2023, pp. 10249--10258.

\bibitem{hu2022coarse}
Z.~Hu, G.~Lu, J.~Guo, S.~Liu, W.~Jiang, D.~Xu, Coarse-to-fine deep video coding with hyperprior-guided mode prediction, in: {Proceedings of the IEEE/CVF Conference on Computer Vision and Pattern Recognition}, 2022, pp. 5921--5930.

\bibitem{tong2022videomae}
Z.~Tong, Y.~Song, J.~Wang, L.~Wang, {VideoMAE}: Masked autoencoders are data-efficient learners for self-supervised video pre-training, in: {Advances in Neural Information Processing Systems}, Vol.~35, 2022, pp. 10078--10093.

\bibitem{fourure2017residual}
D.~Fourure, R.~Emonet, E.~Fromont, D.~Muselet, A.~Tremeau, C.~Wolf, Residual conv-deconv grid network for semantic segmentation, in: {Proceedings of the British Machine Vision Conference}, 2017, pp. 1--13.

\bibitem{zamir2022restormer}
S.~W. Zamir, A.~Arora, S.~Khan, M.~Hayat, F.~S. Khan, M.-H. Yang, Restormer: Efficient transformer for high-resolution image restoration, in: {Proceedings of the IEEE/CVF Conference on Computer Vision and Pattern Recognition}, 2022, pp. 5728--5739.

\bibitem{wang2004image}
Z.~Wang, A.~C. Bovik, H.~R. Sheikh, E.~P. Simoncelli, Image quality assessment: from error visibility to structural similarity, {IEEE Trans. on Image Process.} 13~(4) (2004) 600--612.

\bibitem{zhang2018unreasonable}
R.~Zhang, P.~Isola, A.~A. Efros, E.~Shechtman, O.~Wang, The unreasonable effectiveness of deep features as a perceptual metric, in: {Proceedings of the IEEE/CVF Conference on Computer Vision and Pattern Recognition}, 2018, pp. 586--595.

\bibitem{xue2019video}
T.~Xue, B.~Chen, J.~Wu, D.~Wei, W.~T. Freeman, Video enhancement with task-oriented flow, {Int. J. Comput. Vis.} 127~(8) (2019) 1106--1125.

\bibitem{mercat2020uvg}
A.~Mercat, M.~Viitanen, J.~Vanne, {UVG} dataset: 50/120fps 4k sequences for video codec analysis and development, in: {Proceedings of the ACM International Conference on Multimedia}, 2020, pp. 297--302.

\bibitem{bossen2013common}
F.~Bossen, et~al., Common test conditions and software reference configurations, JCTVC-L1100 12~(7) (2013).

\bibitem{hm}
HM-17.0, \url{ https://vcgit.hhi.fraunhofer.de/jvet/HM/}, accessed 2023-03-01 (2023).

\bibitem{vtm}
VTM-19.0, \url{ https://vcgit.hhi.fraunhofer.de/jvet/VVCSoftware_VTM/}, accessed 2023-03-01 (2023).

\bibitem{yilmaz2021end}
M.~A. Y{\i}lmaz, A.~M. Tekalp, End-to-end rate-distortion optimized learned hierarchical bi-directional video compression, {IEEE Trans. on Image Process.} 31 (2021) 974--983.

\bibitem{VCEG-M33}
G.~Bjontegaard, Calculation of average {PSNR} differences between {RD}-curves., VCEG-M33 (2001).

\end{thebibliography}

\end{document}